\documentclass[aps,showpacs,
,pra
]{revtex4}
\usepackage{amsmath}
\usepackage{color}
\usepackage{supercite}
\usepackage{graphicx}
\usepackage{amssymb}

\def\ket#1{|#1\rangle}

\begin{document}

\title{Rotating Fermi gases in an anharmonic trap}

\author{Kiel Howe}
\email{howek@email.arizona.edu}
\affiliation{Department of Physics, University of Arizona\\Tucson, AZ 85721, USA}
\author{Aristeu R. P. Lima}
\email{lima@physik.fu-berlin.de}
\affiliation{Institut f\"{u}r Theoretische Physik, Freie Universit\"{a}t Berlin, Arnimallee 14, 14195 Berlin, Germany}
\author{Axel Pelster}
\email{axel.pelster@uni-duisburg-essen.de}
\affiliation{Fachbereich Physik, Universit\"{a}t Duisburg-Essen, Lotharstrasse 1, 47048 Duisburg, Germany}
\affiliation{Institut f\"{u}r Theoretische Physik, Freie Universit\"{a}t Berlin, Arnimallee 14, 14195 Berlin, Germany}


\begin{abstract}
Motivated by recent experiments on rotating Bose-Einstein condensates, we investigate a rotating, polarized Fermi gas trapped in an anharmonic potential. We apply a semiclassical expansion of the density of states in order to determine how the thermodynamic properties depend on the rotation frequency. The accuracy of the semiclassical approximation is tested and shown to be sufficient for describing typical experiments. At zero temperature, rotating the gas above a given frequency $\Omega_{\rm DO}$ leads to a `donut'-shaped cloud which is analogous to the hole found in two-dimensional Bose-Einstein condensates. The free expansion of the gas after suddenly turning off the trap is considered and characterized by the time and rotation frequency dependence of the aspect ratio. Temperature effects are also taken into account and both low- and high-temperature expansions are presented for the relevant thermodynamical quantities. In the high-temperature regime a virial theorem approach is used to study the delicate interplay between rotation and anharmonicity.
\end{abstract}

\maketitle

%
%

\section{Introduction}
\label{intro}
In a remarkable series of experiments in the nineties, sophisticated cooling techniques were used to expose the quantum nature of ultracold dilute gases. For bosons, the realization of Bose-Einstein condensation (BEC) in 1995 \cite{mh.anderson-s269,PhysRevLett.75.3969,PhysRevLett.78.985} revealed the importance of the underlying aggregating statistics for the existence of a broken symmetry phase in weakly interacting systems. For fermions, the statistics has just the opposite effect. Due to the Pauli exclusion principle, low energy collisions of fermionic particles are strongly suppressed, worsening the thermalization between the atoms and, therefore, making the cooling techniques more involved. Nevertheless, a few years later another milestone experiment was realized in which a quantum degenerate Fermi gas was observed \cite{deMarco-s285}. Since then, research with both types of degenerate quantum gases has experienced a strong growth, and different trends have been pursued, each showing interesting and relevant connections to other fields of physics such as atomic, molecular and condensed matter physics.

Rotating quantum gases constitute one of these fast growing areas. In the case of bosons in a harmonic trap, intensive investigations have been carried out, both theoretically \cite{RevModPhys.71.463,RevModPhys.73.307,pethick,pitaevskii,stringari-review,bloch:885,fetter-rmp} and experimentally \cite{PhysRevLett.92.050403,EurophysLett65,LaserPhysLett2}. If the rotation frequency $\Omega$ is of the order of (but smaller than) the radial trapping frequency $\omega_{\perp}$, i.e. $\Omega\lesssim\omega_{\perp}$, the rotation is called slow. In this regime, intrinsic properties of rotating systems like the temperature dependence of the moment of inertia \cite{PhysRevLett.76.1405}, rotation induced excitations \cite{PhysRevLett.83.4452}, and the formation of vortices \cite{PhysRevLett.84.806,PhysRevLett.83.2498} have been considered. If one rotates faster, vortex arrays are created \cite{Abo-Shaeer-s292} where small-amplitude oscillations, so called Tkachenko modes, have been observed \cite{PhysRevLett.91.100402} and theoretically investigated \cite{PhysRevA.65.063611,PhysRevLett.92.160405,PhysRevLett.91.110402,sonin:011603}. In addition, fast rotating ($\Omega\rightarrow\omega_{\perp}$) BECs have been linked to fascinating many-body phenomena such as the integer quantum Hall effect \cite{PhysRevLett.87.060403,PhysRevLett.90.140402} or its fractional version \cite{PhysRevLett.87.120405}, and are suspected to display quantum phase transitions to non superfluid strongly correlated states \cite{PhysRevLett.84.6,PhysRevLett.80.2265}.

In the fast rotation regime, experiments in harmonic traps are limited by the loss of confinement due to centrifugal forces when the rotation frequency approaches the axial harmonic trapping frequency. To overcome this problem, A. L. Fetter suggested adding a quartic term to the trap potential \cite{PhysRevA.64.063608}. The successful experimental implementation of such a quartic term was carried out in the group of J. Dalibard in Paris \cite{PhysRevLett.92.050403}. In their set-up a laser is used to stir the quartically trapped condensates, making possible the exploration of many properties of the system such as, for instance, the vortex distribution, the optical density \cite{PhysRevLett.92.050403}, and the shape oscillations \cite{EurophysLett65}. The experimental realization triggered further theoretical research focusing on Bose-Einstein condensates (BEC). For instance, their collective oscillations \cite{cozzini:100402}, thermodynamical \cite{kling:023609} and dynamical \cite{kling2} properties have been considered in the presence of the quartic term.

Thus, the question arises, what new features come about through the interplay between rotation and fer\-mionic statistics. In a harmonic trap, some properties have already been investigated. For instance, the moment of inertia and its relation to quadrupole oscillations, in a framework valid for bosons and fermions \cite{PhysRevA.63.033602}, and the collective excitations \cite{PhysRevLett.91.070401} have been studied. In the superfluid phase, formation of vortices \cite{zwierlein-sc435} has been observed and a rotation-induced phase separation has been proposed \cite{unitary}. In addition, the influence of the Landau levels on the density profiles of two-dimensional polarized fermions in the context of the integer quantum Hall effect \cite{PhysRevLett.85.4648} and the experimental feasibility of fractional quantum Hall states \cite{baranov:070404} have also been theoretically considered.

In the case of an anharmonic trap, apart from studying the breathing mode in the Bardeen-Cooper-Schrieffer (BCS)-BEC crossover regime \cite{silva:023623}, fermions have not been considered and most of the interesting aspects of the phy\-sics of rotating systems remain restricted to pure harmonic confinement. At low temperatures, polarized atomic Fermi gases are essentially non-interacting due to Pauli blocking and the fact that p-wave interactions are negligible in such systems. However, due to the Fermi pressure, particles tend to avoid the center of the trap, so that, effectively, a statistical interaction arises even in the case of an ideal gas, which bears some analogies with repulsively interacting bosons \cite{PhysRevA.55.4346}.

In this work we are concerned with rotating ideal Fermi gases of spin polarized particles rotating with frequency $\Omega$. We consider the same trap configuration as in the Paris experiment, which will be described in detail in Section \ref{sec:2}. In Section \ref{sec:3} we outline a general approach to formally justify the semiclassical approximation for calculating the density of states in anharmonic traps. Furthermore, we apply this technique to the Paris trap and see that a treatment within the semiclassical approximation is justified. In Section \ref{sec:4} we consider ground-state properties of such a system in a similar semiclassical approach. We start by evaluating the Fermi energy for arbitrary rotation, then we obtain the particle density, the momentum distribution, and free expansion time-of-flight absorption pictures. In Section \ref{sec:5} temperature effects are considered and we derive expressions for the grand-canonical thermodynamic properties in terms of incomplete Fermi functions. In addition we analyze analytic low- and high-temperature expansions. For the latter, a virial theorem approach is also presented. Section \ref{sec:6} discusses the main results of this work emphasizing the effects of rotation and anharmonicity in comparison to a non-rotating harmonic trap and presents an outlook for the field.

\section{Anharmonic Trap Potential}
\label{sec:2}

In the following we consider a trap potential which is anharmonic and consists of harmonic axially symmetric magnetic
trapping supplemented by an additional Gaussian trapping term in the $xy$-plane. The Gaussian laser beam is approximated by its fourth order Taylor series
\begin{equation}
U({r_{\perp}}) \approx U_{0} - \dfrac{2U_{0}}{w^{2}}r_{\perp}^{2}+ \dfrac{2U_{0}}{w^{4}}r_{\perp}^{4},\label{eq:pot-2_las}
\end{equation}
where $U_{0}$ and $w$ are the laser intensity and width, respectively. The radial distance to the origin is denoted by $r_{\perp}=\sqrt{x^{2}+y^{2}}$. The first term in \eqref{eq:pot-2_las} is just a shift in the energy scale and will be neglected in the following. The second order term will be absorbed into the harmonic trapping term. The last term is the one responsible for the quartic confinement and is characterized by the anharmonicity constant $k={2U_{0}}/{w^{4}}$. It is important to notice that the harmonic trapping in the $z$-direction is not affected by the Gaussian laser potential and is characterized by its harmonic oscillator frequency $\omega_{z}$.

An issue in the field of rotating quantum gases is that the rotation frequency, with which the condensate effectively rotates, might be different from the one of the stirring laser, due to changes in the shape dependent rotation properties of the gas such as the moment of inertia \cite{LaserPhysLett2}. We neglect these effects and assume that a stirring laser can adiabatically add angular momentum to the particles, so that the gas rotates with frequency $\Omega$, yielding an additional centrifugal harmonic term to the trap potential in the co-rotating frame \cite{pethick,pitaevskii}. Under these considerations, the rotating anharmonic potential takes the form \cite{kling:023609}
\begin{equation}
V({\bf x}) = \dfrac{\epsilon_{z}}{2} \left( \lambda^{2} \eta \dfrac{r_{\perp}^{2}} {a_{z}^{2}} + \dfrac{z^{2}}{a_{z}^{2}} + \dfrac{\kappa}{2} \dfrac{r_{\perp}^{4}}{a_{z}^{4}}\right),\label{eq:pot-2}
\end{equation}
where $\epsilon_{z}=\hbar\omega_{z}$ and $a_{z}=\sqrt{\hbar/M\omega_{z}}$ are the axial harmonic energy and length scales, respectively, $\lambda=\omega_{\perp}/\omega_{z}$ denotes the anisotropy of the trap, $\kappa=ka^{4}_{z}/\epsilon_{z}$ represents the anharmonicity of the trap, and $M$ is the mass of the atomic species under consideration. The rotation is taken into account with the parameter $\eta=1-\Omega^{2}/\omega_{\perp}^{2}$. Adopting the values given for the Paris experiment \cite{PhysRevLett.92.050403}, the resulting rotating trap is then described by the radial harmonic trapping frequency $\omega_{\perp}=2\pi\times64.8(3)\,{\rm Hz}$, the axial harmonic trapping frequency $\omega_{z}=2\pi\times11.0\,{\rm Hz}$, the force constant $k=2.6(3)\times10^{-11}\,{\rm Jm}^{-4}$ of the quartic trapping term, the anisotropy $\lambda\approx6$, and a tunable rotation frequency $\Omega$ induced by the stirring laser.

In the Paris experiment, a BEC was obtained with the bosonic isotope $\,^{87}{\rm Rb}$, for which the dimensionless anharmonicity takes the value $\kappa\approx0.4$. Since fermionic species like $\,^{40}{\rm K}$ and $\,^{6}{\rm Li}$ have been sympathetic cooled by using the bosonic atoms $\,^{87}{\rm Rb}$ \cite{ospelkaus:020401,PhysRevLett.89.150403} and $\,^{23}{\rm Na}$ \cite{PhysRevLett.88.160401}, respectively, we consider them likely candidates for experiments with ultracold degenerate Fermi gases in the Paris trap. For definiteness, we evaluate all results of this paper for the parameters of the trap of the Paris experiment and the $\,^{40}{\rm K}$ species, for which we have $\kappa\approx1.9$ (see Table \ref{tab:1}).

\begin{figure}
\begin{centering}
\includegraphics[width=3in]{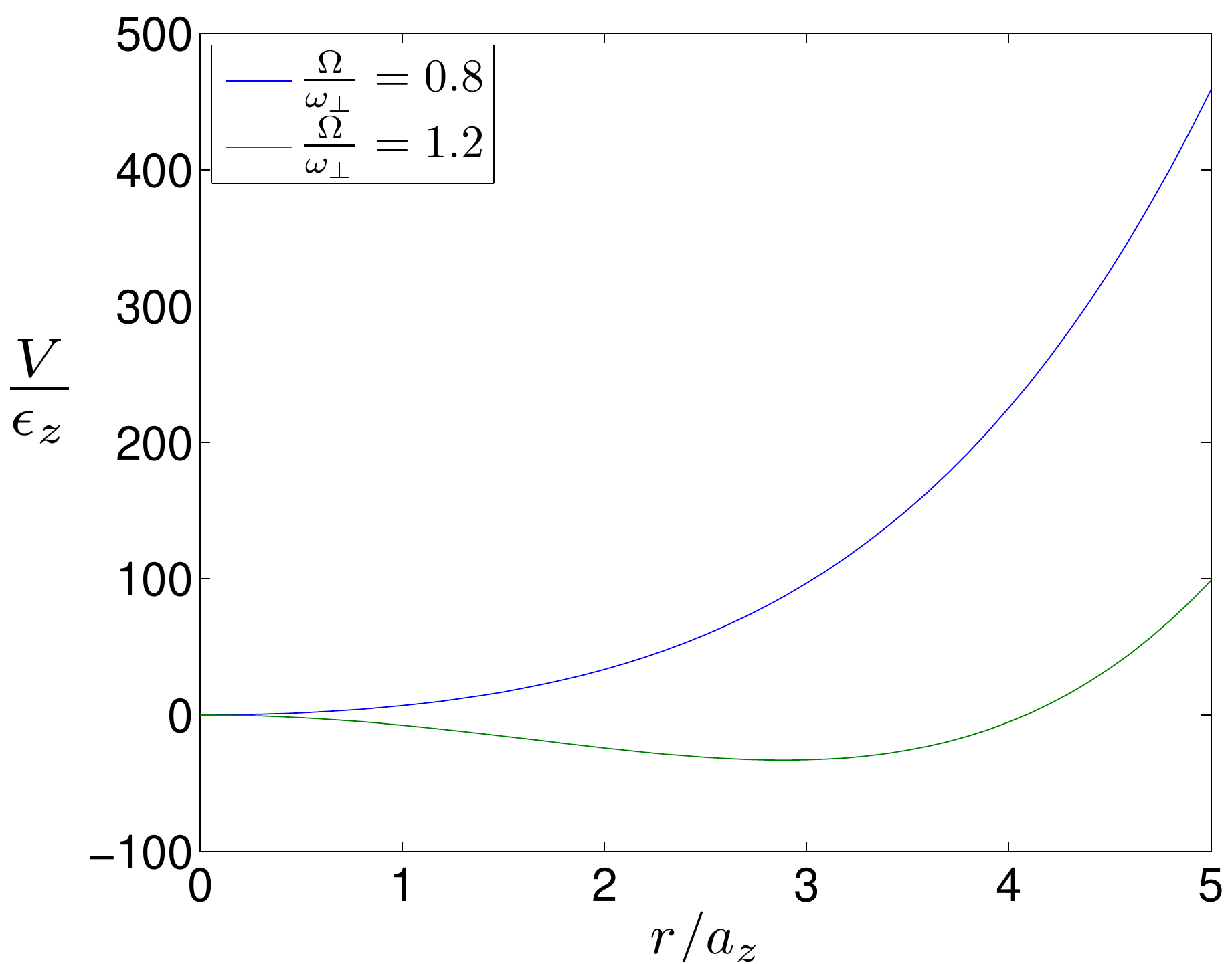}
\par\end{centering}
\caption{(Color online) The anharmonic potential \eqref{eq:pot-2} as a function of radial position in the plane $z=0$, for the parameters given at Table \ref{tab:1} at both sub- and supercritical rotations. For the latter, a minimum at $\tilde{r}_{\perp}\neq0$ appears due to the centrifugal nature of the harmonic term.\label{fig:1}}       
\end{figure}

Let us discuss the trap potential \eqref{eq:pot-2} and some of its peculiarities concerning rotation. For the moment, we restrict ourselves to the $xy$-plane. As depicted in Fig.~\ref{fig:1}, for slow rotations, i.e. $\Omega < \omega_{ \perp} (\eta>0)$, the minimum of the potential is located at the origin. Rotating faster one eventually reaches the point where the harmonic confinement vanishes ($\Omega=\omega_{\perp}, \eta=0$) and the trap is purely quartic. Hereafter this will be called critical rotation. If the rotation is made even faster, one achieves the supercritical regime, where the harmonic confinement becomes centrifugal and the trap minimum is moved to $\tilde{r}_{\perp} = \sqrt{-\eta\lambda^{2} a_{z}^{2} /\kappa}$ and, with respect to the origin, has a depth given by
\begin{equation}
\Delta E = \lambda^{4}\eta^{2}\epsilon_{z}/4\kappa,
\label{eq:DeltaE}
\end{equation}
which naturally characterizes the anharmonic trap. The lowest available potential energy is, thus, $-\Delta E$, in the supercritical case.

\begin{table}
\caption{Experimental values of the parameters of the anharmonic trap considered in this work.}
\label{tab:1}       
\begin{tabular}{llllll}
\hline\noalign{\smallskip}
$\frac{\omega_{\perp}}{2\pi}$ (Hz) & $\frac{\omega_{z}}{2\pi}$ (Hz) &  $ \lambda $    &  $k $ (Jm$^{-4}$) & $\kappa $ & $N$ \\
\noalign{\smallskip}\hline\noalign{\smallskip}
$ 64.8(3)$ & $ 11.0$   & $ 6$ & $2.6(3)\times 10^{-11}$& $ 1.9$ & $3\times 10^{5}$  \\
\noalign{\smallskip}\hline
\end{tabular}
\end{table}

\section{Semiclassical Approximation for Density of States}
\label{sec:3}
In this section we briefly discuss the meaning and the implementation of the semiclassical approximation for a system of particles possessing one-body Hamiltonian $H$ with the eigenvalue equation
\begin{equation}
H\ket{\psi_{n}} = E_{n}\ket{\psi_{n}},
\end{equation}
where $\ket{\psi_{n}}$ and $E_{n}$ denote the $n$-th single-particle eigenstate and eigenvalue, respectively.

In view of studying the various thermodynamic properties of the system, one is usually interested in evaluating sums of the form
\begin{equation}
\sum_{n}F(E_{n}),
\end{equation}
where $F$ is some function of the single-particle energy levels. The semiclassical approximation consists in considering the energy levels so close to one another that the sum over them can be taken as an integral over a phase space density:
\begin{equation}
\sum_{n}F(E_{n})=\int_{}g(E)F(E)dE.
\label{semi_sum}
\end{equation}
To this end we define the density of states
\begin{equation}
g(E)=\sum_{n}\delta(E-E_{n}),\label{eq:g-1a}
\end{equation}
where $\delta$ is the Dirac delta distribution. In the following we investigate for the parameters of the Table \ref{tab:1} whether it is justified to approximate the density of states semiclassically according to
\begin{equation}
g_{0}(E)=\int\dfrac{d^{D}xd^{D}p}{(2\pi\hbar)^{D}}\,\delta[E-H_{}({\mathbf x},{\mathbf p})],\label{eq:g-1b}
\end{equation}
where $H_{}({\mathbf x},{\mathbf p})={{\mathbf p} }^{2}/(2M)+V({\mathbf x})$ denotes the single-particle classical Hamiltonian. In systems
where the exact energy spectrum is known, e.g. a harmonic trap, the validity of this approximation can be determined by comparing \eqref{eq:g-1a} and \eqref{eq:g-1b}. However, in the case that the energy spectrum is unknown, as for the present anharmonic potential \eqref{eq:pot-2}, we must take a different approach to
justify the semiclassical approximation. To this end, we determine not only the leading semiclassical approximation $g_{0}(E)$, but also its subleading correction $g_{1}(E)$. After setting the relevant energy scale of the system, the validity of the semiclassical approximation can be determined by the {\sl relative} magnitude of the correction. We first treat the general case and, then, specialize to the anharmonic trap of the Paris experiment.

\subsection{Semiclassical Expansion}
\label{sec:3A}

To obtain the semiclassical expansion of the density of states of the anharmonic potential, we follow the general procedure worked out in Ref. \cite{kleinert-pi}, which derives the density of states in form of a gradient expansion
\begin{equation}
g(E) = g_{0}(E) + g_{1}(E) + \cdots.
\label{eq:approxZero}
\end{equation}
The first term corresponds to the semiclassical approximation $g_{0}(E)$
in \eqref{eq:g-1b}, which yields after integration over the momenta
\begin{equation}
g_{0}(E)=\left(\dfrac{M}{2\pi\hbar_{}^{2}}\right)^{\frac{D}{2}}\int d^{D}x\, \Theta\left[E-V({\mathbf x})\right]\dfrac{\left[E-V({\bf x})\right]^{\frac{D}{2}-1}}{\Gamma(D/2)}.\label{eq:approx}
\end{equation}
The second term is the subleading correction to the density of states
\begin{eqnarray}
g_{1}(E) & = & -\dfrac{\hbar_{}^{2}}{24M} \left[\dfrac{M}{2\pi\hbar_{}^{2}} \right]^{\frac{D}{2}}\dfrac{1}{\Gamma(D/2-2)}\nonumber\\
& & \times\int d^{D}x\,\Theta\left[E-V({\bf x})\right]\nabla^{2}V({\bf x})\left[E-V({\bf x})\right]^{\frac{D}{2}-3}.\nonumber\\
\label{eq:g1}
\end{eqnarray}

We note that, for the physically relevant dimension $D=3$, the spatial integral
\eqref{eq:g1} diverges due to the behavior of the term $\left[E-V({\bf x})\right]^{\frac{D}{2}-3}$
at the classical turning points. This problem can be solved through
the method of analytic continuation as demonstrated later by applying
\eqref{eq:g1} to the anharmonic potential \eqref{eq:pot-2}.

It remains to determine the energy scales for which the semiclassical approach is supposed to be valid. At high enough temperatures, the particles are classical. When the temperature is gradually lowered, or other parameters are changed, classical statistical mechanics is no longer exact. At first, a semiclassical approach may work, but for certain conditions it may completely fail. The temperature, particle number, and trap potential parameter scales in which the semiclassical approximation is valid depend on the underlying statistics. For Bosons in the gas phase, the thermal distribution is characterized by the thermal energy $k_{B}T$. If this energy is sufficiently larger than the the energy of the single-particle ground-state, the leading semiclassical approximation is valid. At the BEC critical temperature $T_{c}$, however, the ground-state occupancy becomes macroscopic. This effect must be treated outside of semiclassics because it corresponds to a broken symmetry state associated to the divergence of the Bose distribution $(e^{\beta E}-1)^{-1}$ for $E=0$. We will see explicitly in the case of the anharmonic oscillator that the subleading correction is not negligible for energies near the ground-state energy. The thermal energy at the critical temperature is thus the relevant energy scale for determining the validity of this approach.

For Fermions the considerations are slightly different. Due to the Pauli exclusion principle, as $T\rightarrow0$ the particles in the system do not all drop to the ground-state but rather fill every energy level up to the Fermi energy $E_{F}$, which is defined as the energy of the highest occupied state. As the density of states $g(E)$ typically increases with a power of $E$, the majority of particles in the system will have energies much larger than the ground-state value and their average energy is of order $E_{F}$, even at $T=0$, provided $N$ is large. Therefore, in principle, a spin-polarized Fermi gas can be well described at all temperatures in the semiclassical approximation provided the particle number is large enough. The relevant energy determining the validity of this assertion is simply the Fermi energy $E_{F}$. Numerical calculations indicate that this is indeed the case \cite{phd}.

\subsection{Leading Semiclassical Approximation}
\label{sec:3B}

To consider the anharmonic potential \eqref{eq:pot-2}, we must first determine explicitly the leading and subleading terms of the semiclassical density of states. We can, then, apply the leading term to calculate the Fermi energy, and with this result return to verify the a priori assumption of the validity of the approximation.

\begin{figure}
\begin{centering}
\includegraphics[width=8.cm]{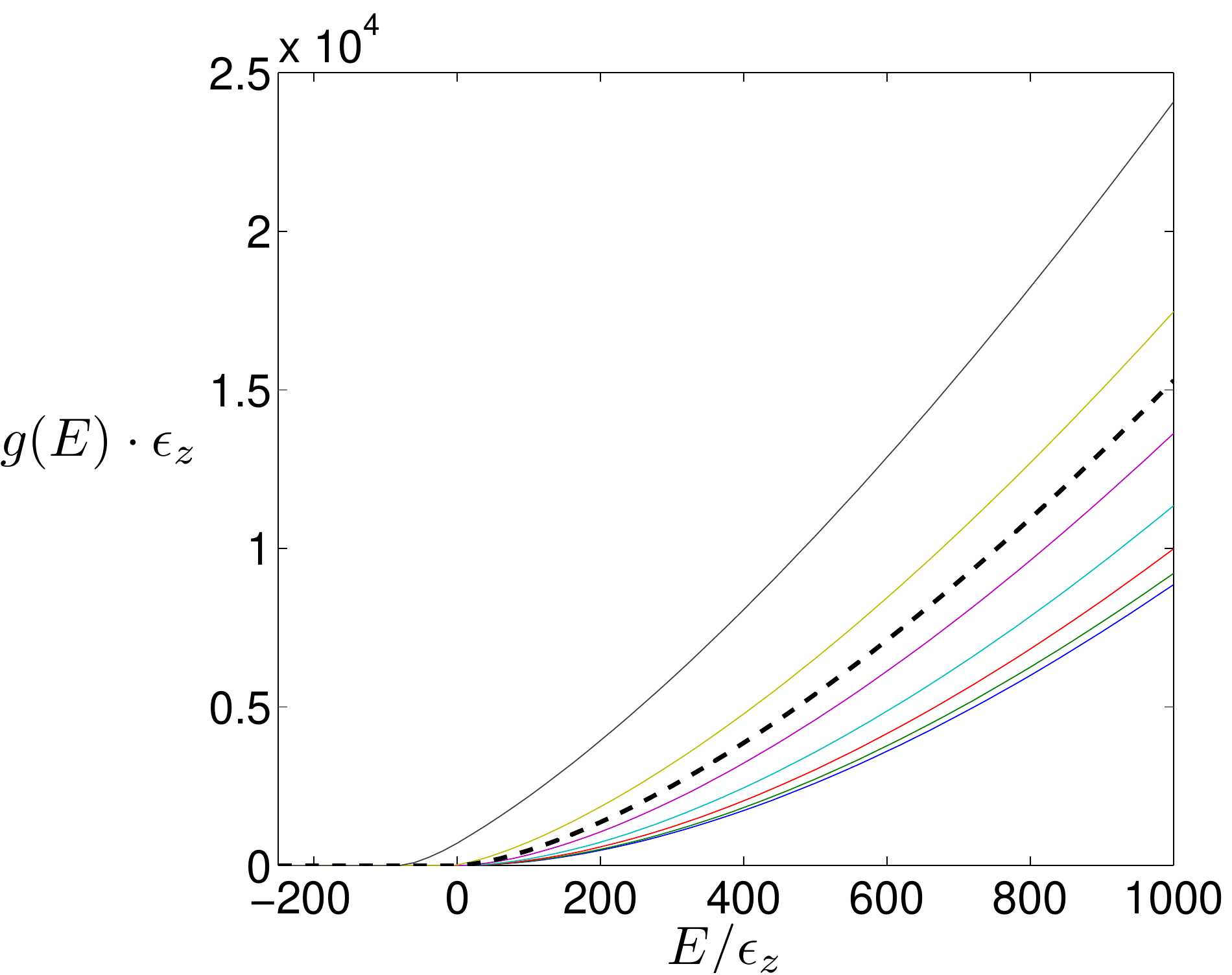}
\par\end{centering}
\caption{(Color online) Density of states \eqref{eq:g-2} with the parameters of Table \ref{tab:1}  for the anisotropy and the anharmonicity of the trap. The solid lines are from bottom to top at rotations of $\Omega=0.1\,\omega_{\perp}$
to $\Omega=1.3\,\omega_{\perp}$ in steps of $\Delta\Omega=0.2\,\omega_{\perp}$.
The dashed line is the density of states at the critical rotation
$\Omega=\omega_{\perp}$.\label{fig:2}}
\end{figure}

The form of the semiclassical density of states in the anharmonic potential \eqref{eq:pot-2} can be found from \eqref{eq:approx}. Performing the spatial integral we obtain the result
\begin{eqnarray}
g_{0}(E) & = & \begin{cases}
\dfrac{2(E + \Delta E)^{\frac{3}{2}}}{3\epsilon_{z}^{\frac{5}{2}}\kappa^{\frac{1}{2}}}-\dfrac{\lambda^{2}\eta E}{2\epsilon_{z}^{2}\kappa}-\dfrac{\lambda^{2}\eta^{}\Delta E}{3\epsilon_{z}^{2}\kappa^{}};& E\geq0\\
\dfrac{4(E + \Delta E)^{\frac{3}{2}}}{3\epsilon_{z}^{\frac{5}{2}}\kappa^{\frac{1}{2}}}; &\!\!\!\!\!\!\!\!\!\!\!\!\!\!\!\!\!\!\! - \Delta E\leq E\leq0\end{cases},\nonumber\\
\label{eq:g-2}
\end{eqnarray}
where it is meant in this and other equations, that the function takes on nonzero values for $E<0$ only for super-critical rotations.
The density of states \eqref{eq:g-2} for the parameters of the anharmonic trap described in Table \ref{tab:1} is shown at several rotation frequencies $\Omega$ in Fig.~\ref{fig:2}. For subcritical rotations, the density of states vanishes for $E<0$. For supercritical rotations, the density of states is nonzero also for $- \Delta E\leq E\leq0$, due to the dip in the potential.

There are two important limiting cases which shed some light on the behavior of the system. The first one is the harmonic limit $\kappa\rightarrow0$, for which we have from \eqref{eq:g-2}
\begin{equation}
g_{0}^{\rm HO}(E)=\dfrac{E^{2}}{2\lambda^{2}\eta\epsilon_{z}^{3}},\label{eq:gHO}
\end{equation}
which corresponds to the known result \cite{RevModPhys.71.463}. Note that the harmonic limit \eqref{eq:gHO} is valid only for subcritical rotations; for supercritical rotations the density of states \eqref{eq:gHO} becomes negative corresponding to an instability due to the fact that particles escape the trap.

The second physically important case is the pure quartic limit in radial confinement. It is realized by either $\eta=0$, i.e. a critical rotation, or $\kappa\gg\lambda^{2}\eta$, so that the density of states \eqref{eq:g-2} takes the form
\begin{equation}
g_{0}^{{\rm QU}}(E)=\dfrac{2E^{\frac{3}{2}}}{3\epsilon_{z}^{\frac{5}{2}}\kappa^{\frac{1}{2}}},\;\;\;\eta=0,\; E>0.\label{eq:gcrit}
\end{equation}
Examining \eqref{eq:DeltaE} and \eqref{eq:g-2}, we see that the rotation parameter $\eta$ appears in the density of states only in ratios to the anharmonicity parameter $\kappa$. For supercritical rotations, we will find that there is a symmetry between the range of rotations, $\eta\rightarrow-\infty$ ($\Omega\rightarrow\infty$) to $\eta=0$ (critical rotation), and the range of anharmonicities from small $\kappa$ to $\kappa\rightarrow\infty$. The supercritical properties of the system can thus be seen as an interpolation between the harmonic limit \eqref{eq:gHO} and the pure quartic limit \eqref{eq:gcrit}.

\subsection{Subleading Semiclassical Approximation}
\label{sec:3C}

We now calculate the subleading semiclassical approximation $g_{1}(E)$ as given in \eqref{eq:g1} for the particular case of the anharmonic potential. A direct integration in \eqref{eq:g1} for $D=3$ yields a divergent result. However, we can rewrite the spatial integral in terms of the Beta function
\begin{equation}
\beta(x,\, y)=\int_{0}^{1}ds\, s^{x-1}\,(1-s)^{y-1},
\end{equation}
for which in Ref.~\cite{gradshteyn-tisp} provides the identity
\begin{equation}
\beta(x,\, y)=\dfrac{\Gamma(x)\Gamma(y)}{\Gamma(x+y)}.
\end{equation}
Thus, the formally divergent integrals can be treated as finite based
on the analytic continuation of the Gamma function through the relationship
$\Gamma(\nu+1)=\nu\Gamma(\nu)$.

To evaluate the integral we need the Laplacian of the anharmonic potential, which reads:
\begin{equation}
\nabla^{2}V({\bf x})=\dfrac{\epsilon_{z}}{a_{z}^{2}}\left(2\lambda^{2}\eta+1+4\kappa\dfrac{r_{\perp}^{2}}{a_{z}^{2}}\right).
\end{equation}
With this and the full potential given in \eqref{eq:pot-2}, we obtain
from \eqref{eq:g1} the first correction to the semiclassical density
of states:
\begin{equation}
g_{1}(E) = \begin{cases}
\dfrac{\left(2\lambda^{2}\eta-1\right)\left(E+\Delta E\right)^{-1/2}}{48\kappa^{\frac{1}{2}}\epsilon_{z}^{\frac{1}{2}}}-\dfrac{1}{6\epsilon_{z}}; & E\geq0\\
\dfrac{\left(2\lambda^{2}\eta-1\right)\left(E+\Delta E\right)^{-1/2}}{24\kappa^{\frac{1}{2}}\epsilon_{z}^{\frac{1}{2}}}; & \!\!\!\!\!\!\!\!\!\!\!\!\!\!\!\!\!\!\!\!\! - \Delta E\leq E<0\end{cases},
\label{eq:first-correction}
\end{equation}
where, again, the lower line is valid only for supercritical rotations.

To determine the effects of the trap parameters on the relevance of
the correction term, we examine the ratio $g_{1}(E)/g_{0}(E)$. We
first see that it is in general reduced in both the sub- and supercritical
regimes as $(E+\Delta E)/\epsilon_{z}$ increases. Likewise, the importance
of the correction term decreases as the anharmonicity $\kappa$ decreases.
As the correction is largest for low energies, the most interesting
case for supercritical rotations is when $E<0$, where we have

\begin{equation}
\dfrac{g_{1}(E)}{g_{0}(E)}=\dfrac{2\lambda^{2}\eta-1}{32}\left[\dfrac{\epsilon_{z}}{E +\Delta E} \right]^{2}, \;\;\;\;-\Delta E<E<0.\label{eq:correction-ratio}
\end{equation}
From \eqref{eq:correction-ratio} we see that for a fixed $E + \Delta E$, the relevance of the correction increases linearly with the rotation parameter $\eta$. Furthermore, we see that the correction \eqref{eq:first-correction} diverges as $E\rightarrow -\Delta E$. However, this divergence is only of the form $(E + \Delta E)^{-1/2}$, and so any interesting integral of the form \eqref{semi_sum} over the density of states will still have a finite contribution from this correction, provided $F(E)$ is not itself divergent. For Bose Systems, $F(E)$ contains a factor of $[\exp{\beta(\mu-E)-1}]^{-1}$. The phase transition to Bose-Einstein condensation happens as $\mu \rightarrow E_{0}$, where $E_{0}$ is the lowest energy eigenvalue. Thus, in the condensate phase, an integral of the form \eqref{semi_sum} has an improper lower limit for $F(E)$ blows up as $E\rightarrow-\Delta E$, since it is the lowest available energy. This problem can be overcome by dividing the gas explicitly into a condensate plus thermal particles.

\begin{figure}
\begin{centering}
\includegraphics[width=8cm]{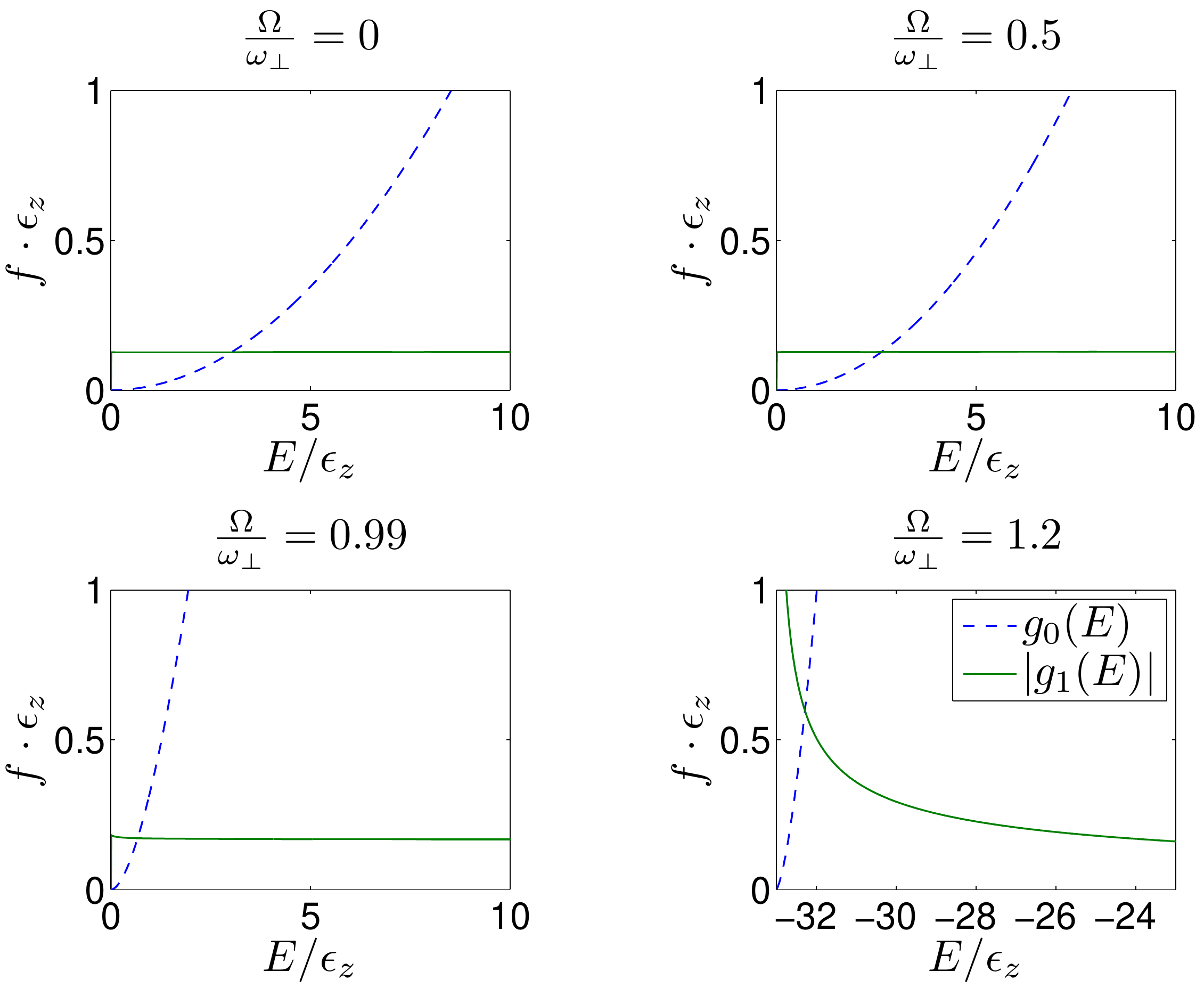}
\par\end{centering}
\caption{(Color online) The semiclassical density of states \eqref{eq:g-2} and the leading
semiclassical correction \eqref{eq:first-correction} for several
rotation frequencies, using the parameters of Table \ref{tab:1}. \label{fig:3}}
\end{figure}

We can now check the validity of the semiclassical approximation in
the case of the anharmonic trap. Both the leading \eqref{eq:g-2}
and the subleading \eqref{eq:first-correction} term are plotted in
Fig.~\ref{fig:3} for the parameters of Table \ref{tab:1} , where we see that even for supercritical rotations the
subleading correction becomes negligible on energy scales of more
than a few $\epsilon_{z}$. In the following section we determine
the Fermi energy $E_{F}$ of the system, from which we will see that, for the
chosen parameters, the difference between $E_{F}$ and the minimum potential is orders of
magnitude larger than $\epsilon_{z}$. With this, we justify the semiclassical approximation \eqref{eq:g-2} to the density of states.

\section{Ground-State Properties}
\label{sec:4}
Though not fully accessible from the experimental point of view, the ground-state and its properties are crucial for understanding the system. The Fermi energy, e.g., carries information about aspects like dimensionality, trap geometry, and particle number and, thus, providing information necessary to handle important physical issues.

In this section, we consider some properties of the ground-state like the dependence of the Fermi energy on the particle number, the ground-state energy, the spatial and momentum distribution and, lastly, the ballistic expansion of the system.

\subsection{Fermi Energy}
\label{sec:4A}

The Fermi distribution is given by
\begin{equation}
f(E,\,\beta,\mu)=\dfrac{1}{e^{\beta(E-\mu)}+1},\label{eq:fermi-distribution}
\end{equation}
where $\mu$ is the chemical potential and $\beta=1/k_{B}T$ with
$k_{B}$ the Boltzmann constant and $T$ the temperature of the system.
At $T=0$, i.e., $\beta\rightarrow\infty$, the Fermi distribution becomes
a step function at the Fermi energy, which is defined as the chemical
potential at $T=0$: \[
\lim_{\beta\rightarrow\infty}f(E,\,\beta,\,\mu)=\Theta(E_{F}-E);\;\;\;\; E_{F}=\left.\mu\right|_{T=0}.\]

The ground-state particle number is therefore given by
\begin{equation}
N(E_{F})=\int_{-\infty}^{E_{F}}g(E)dE.
\end{equation}
Evaluating this integral with the semiclassical density of states from \eqref{eq:g-2} yields
\begin{equation}
N(E_{F})  =  \dfrac{4(E_{F} + \Delta E)^{\frac{5}{2}}} {15\epsilon_{z}^{\frac{5}{2}} \kappa^{\frac{1}{2}}} - \dfrac{\lambda^{2}\eta}{\kappa} \left[ \dfrac{E_{F}^{2}} {4\epsilon_{z}^{2}} +\dfrac{\Delta EE_{F}}{3\epsilon_{z}^{2}}+\dfrac{2{\Delta E}^{2}}{15\epsilon_{z}^{2}}\right]
\label{eq:N-1A}
\end{equation}
for the first case, in which one has $E_{F}\geq0$. For the second case, in which $-\Delta E\leq E_{F}\leq0$, the particle number is given by
\begin{equation}
N(E_{F})= \dfrac{8(E_{F}+\Delta E)^{\frac{5}{2}}}{15\epsilon_{z}^{\frac{5}{2}}\kappa^{\frac{1}{2}}}.\label{eq:N-1B}
\end{equation}
The inverse of (\ref{eq:N-1A}-\ref{eq:N-1B}) determines the Fermi energy as a function of the trap parameters and the particle number. In the harmonic \cite{RevModPhys.71.463} and pure quartic limits we have
\begin{equation}
E_{F}^{\rm HO}=\left(6N\lambda^{2}\eta\right)^{\frac{1}{3}}\epsilon_{z},~ E_{F}^{{\rm QU}}=\left(\dfrac{15}{4}\kappa^{\frac{1}{2}}N\right)^{\frac{2}{5}}\epsilon_{z},\label{eq:EfLimiting}
\end{equation}
respectively. We note that the Fermi temperature $E_{F}^{{\rm QU}}/k_{B}$ is of the same order of magnitude as the critical temperature for non-interacting Bose-Einstein condensates in the quartic regime \cite{LaserPhysLett2}
\begin{equation}
k_{B}T_{C}^{\rm QU} = \epsilon_{z}\left[\frac{2{\kappa}^{\frac{1}{2}}N} {\zeta(5/2){\pi}^{\frac{1}{2}}}\right]^{\frac{2}{5}},
\label{eq:EfLimiting_BEC}
\end{equation}
with $\zeta(5/2)\approx1.342$.

It should be pointed out that negative Fermi energies are allowed by (\ref{eq:N-1A}-\ref{eq:N-1B}) for supercritical rotations,
in which case the gas is confined entirely to the dip in the potential, creating a hole in the center of the trap analogous to the one found in two-dimensional BECs \cite{fetter:013605}.
In this interesting regime ($-\Delta E\leq E_{F}\leq0$)
we see that the Fermi energy also takes on a simple analytic form,
\begin{equation}
E_{F}^{\rm DO}=-\Delta E + \epsilon_{z}\left(\dfrac{15}{8}\kappa^{\frac{1}{2}}N\right)^{\frac{2}{5}},
\label{eq:EfDonut}
\end{equation}
where the superscript DO stands for donut, in reference to the shape that the gas assumes when all the particles are in this energy interval, as discussed below. Notice that in the `donut' regime, for a fixed particle number the Fermi energy is set a distance above the trap minimum that is independent of the rotation frequency, but is directly related to the anharmonicity and to the number of particles.

\begin{figure*}
\begin{centering}
\includegraphics[width=7.cm]{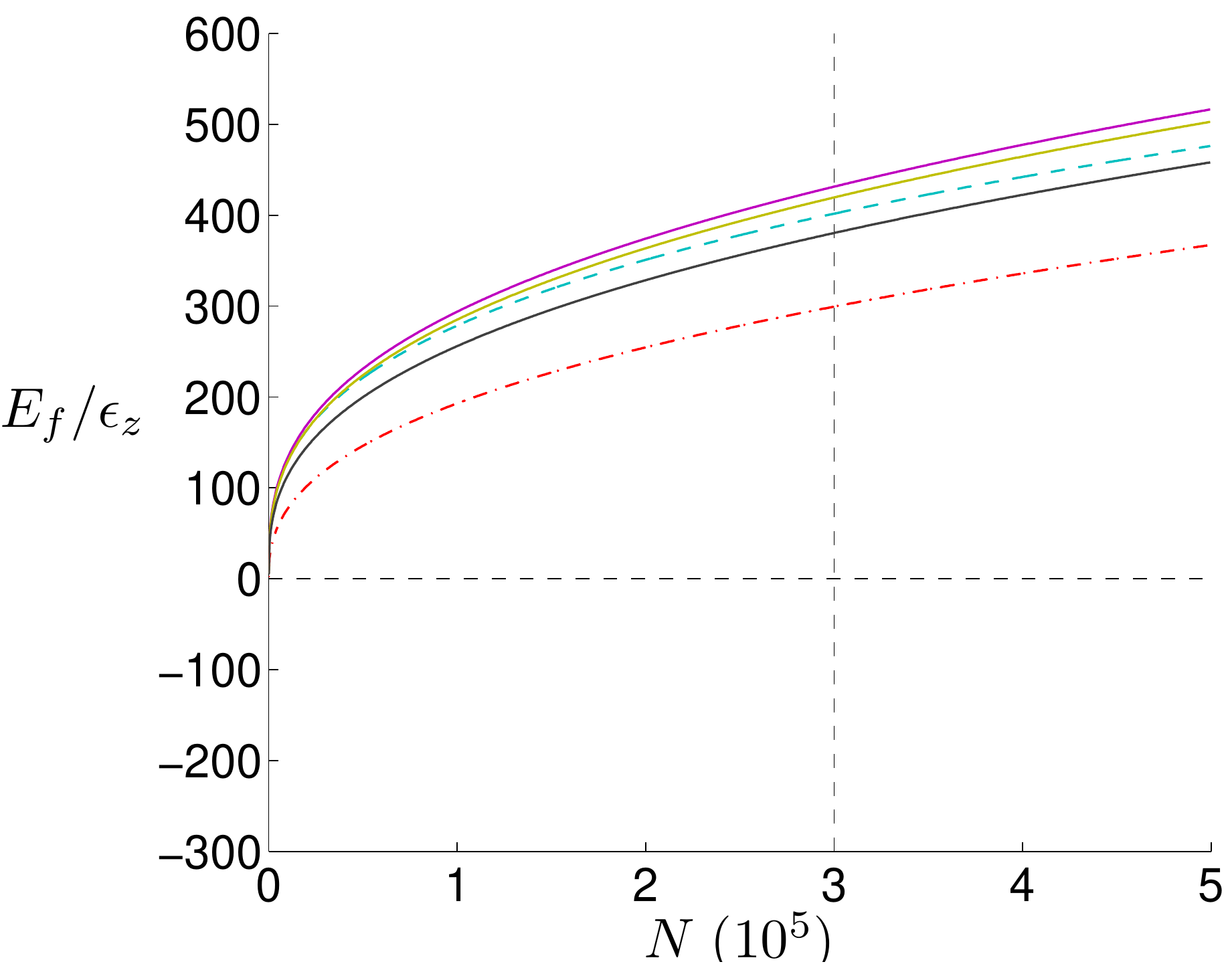}
\hspace{1cm}
\includegraphics[width=7.cm]{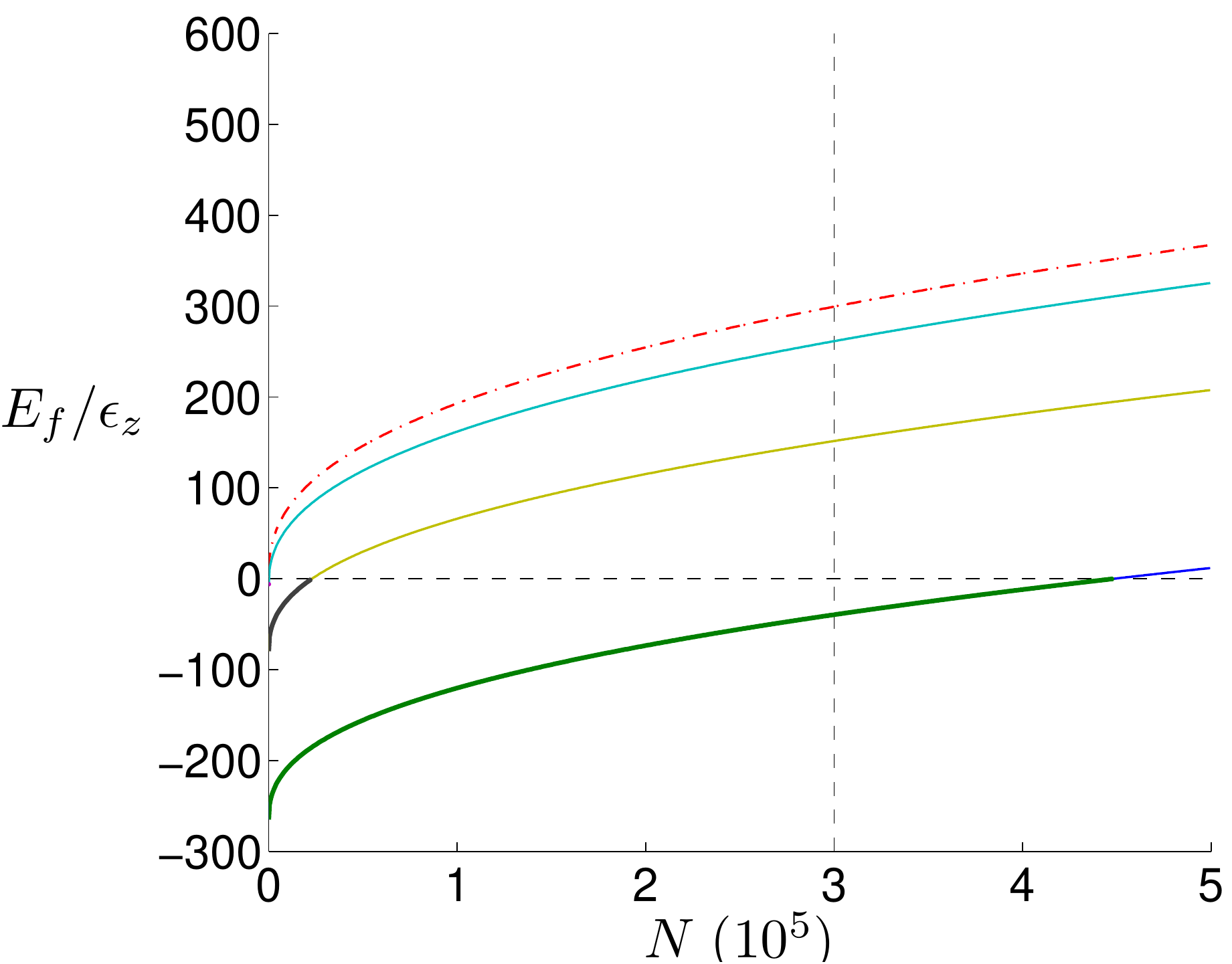}
\par\end{centering}
\caption{(Color online) Relation between Fermi energy and particle number given by (\ref{eq:N-1A}-\ref{eq:N-1B}).
The solid curves are for subcritical rotations $\Omega/\omega_{\perp}=0,\,{1}/{3},\,{2}/{3}$
from top to bottom on the left side, and for supercritical rotations
$\Omega/\omega_{\perp}=1.1,\,1.3,\,1.5$ from top to bottom on the
right side. The dashed curve on the left is the harmonic limit \eqref{eq:EfLimiting}
at zero rotation. The dashed-dotted curve on both sides is the pure
quartic/critical rotation limit \eqref{eq:EfLimiting}. The Fermi
energies in the `donut' regime are given by \eqref{eq:EfDonut} and
are bold-faced. The parameters of the Table \ref{tab:1} were used, and
a guide-line at $N=3\times10^{5}$ indicates the  chosen particle number.\label{fig:4}}
\end{figure*}

We now have three cases where the Fermi energy takes a simple analytic form corresponding to the three interesting rotation regimes: the harmonic limit for subcritical rotations, the pure quartic limit for critical rotations, and the `donut' regime for supercritical rotations. Although we can not obtain from (\ref{eq:N-1A}-\ref{eq:N-1B}) a general explicit expression for the Fermi energy, we can determine its behavior from these special cases. In Fig.~\ref{fig:4} we see that for low $E_{F}$, the general supercritical behavior approaches the harmonic oscillator behavior. As rotation increases to the critical value, the curves approach the pure quartic limit. As the rotation becomes supercritical, the curves are offset to more negative energies and the `donut' regime eventually sets up.

In Fig.~\ref{fig:4} we also see that for any Fermi energy possible, the quantum correction
term to the semiclassical approximation \eqref{eq:first-correction}
is negligible, thus justifying our a priori use of the semiclassical
approximation.

In the following we use the Fermi energy at zero rotation as a convenient energy scale. For our parameters it is\[
E_{F0}\approx430\epsilon_{z}.\]
The Fermi temperature $T_{F}=E_{F}/k_{B}$ at zero rotation provides a corresponding temperature scale, 
\begin{equation}
T{}_{F0}=E_{F}/k_{B}\approx230\,{\rm nK}.
\end{equation}

\subsection{Internal Energy}
\label{sec:4B}
The total internal energy of the system in the ground-state is given
by
\begin{equation}
U(E_{F})=\int_{-\infty}^{E_{F}}Eg(E)dE.
\end{equation}
Evaluating this integral with the density of states from \eqref{eq:g-2}
yields
\begin{equation}
U(E_{F}) =  \begin{cases}
\dfrac{W(E_{F})}{2} - \dfrac{\lambda^{2}\eta}{\kappa}\left[ \dfrac{E_{F}^{3}} {\epsilon_{z}^{2}} + \dfrac{\Delta EE_{F}^{2}}{\epsilon_{z}^{2}} - \dfrac{8{\Delta E}^{3}} {35\epsilon_{z}^{2}}\right]\\
W(E_{F}) \end{cases}
\label{eq:U-1}
\end{equation}
with the remark that the upper line is valid for $E_{F}\geq0$ and the lower one for $- \Delta E\leq E_{F}\leq0$ and the abbreviation
\begin{equation}
W(E_{F}) = \dfrac{8E_{F}\left(E_{F}+\Delta E\right)^{\frac{5}{2}}} {15\epsilon_{z}^{\frac{5}{2}}\kappa^{\frac{1}{2}}} -\dfrac{16\left(E_{F}+\Delta E\right)^{\frac{7}{2}}}{105\epsilon_{z}^{\frac{5}{2}}\kappa^{\frac{1}{2}}}.\label{eq:U-1prime}
\end{equation}
In the harmonic and purely quartic limits, respectively, the internal energy it is related to the particle number and Fermi energy by
\begin{equation}
U^{\rm HO}=\dfrac{3}{4}NE_{F}^{\rm HO},\;\;\;\; U^{{\rm QU}}=\dfrac{5}{7}NE_{F}^{{\rm QU}}.\label{eq:Ulimiting}
\end{equation}
There is also a simple relation in the `donut' regime,
\begin{equation}
U^{\rm DO}=\left(\dfrac{5}{7}E_{F}^{\rm DO}-\dfrac{2}{7}\Delta E\right)N.\label{eq:Udonut}
\end{equation}

\begin{figure}
\begin{centering}
\includegraphics[width=8cm]{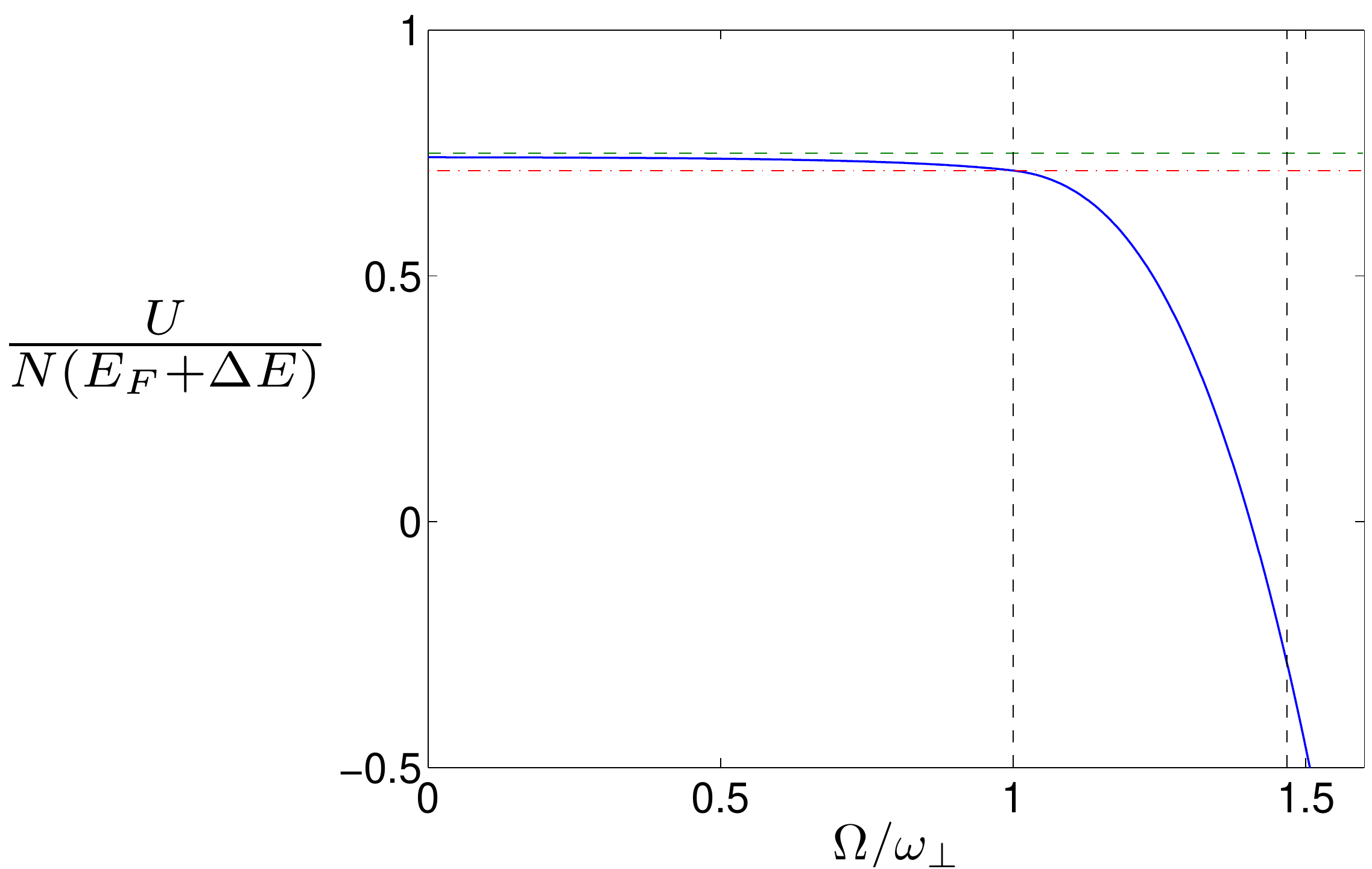}
\par\end{centering}
\caption{(Color online) Ratio of internal energy \eqref{eq:U-1} per particle to $E_{F}+\Delta E$
for fixed $N$ and the parameters
of Table \ref{tab:1}. The dashed horizontal line is the harmonic
limit and the dot-dashed horizontal line is the quartic limit \eqref{eq:Ulimiting}.
The vertical lines indicate the critical rotation ($\Omega = \omega_{\perp}$) and the start of
the `donut' regime \eqref{eq:Udonut}, where $E_{F}<0$.\label{fig:5}}

\end{figure}

We see in Fig.~\ref{fig:5} that the internal energy \eqref{eq:U-1} for subcritical rotations is bounded by the $\kappa\rightarrow0$ harmonic oscillator limit and the $\kappa\rightarrow\infty$ pure quartic limit \eqref{eq:Ulimiting}. This is due to the symmetry between the anharmonicity and the rotation, which causes the latter to enhance the former.

\subsection{Particle Density \label{part_dens}}
\label{sec:4C}

To obtain the particle density and the momentum distribution in the semiclassical approximation, we start from the Wigner function \cite{wigner} in the semiclassical approximation
\begin{equation}
\nu({\mathbf x},{\mathbf p}) = \frac{1}{1+\exp\{\beta[H({\mathbf x},{\mathbf p})-\mu]\}},
\label{wig_func}
\end{equation}
which delivers the particle density through
\begin{equation}
n({\mathbf x}) = \int \frac{d^{3}p}{(2\pi\hbar)^{3}}\nu({\mathbf x},{\mathbf p})
\label{wig_dens}
\end{equation}
and the momentum distribution through
\begin{equation}
n({\mathbf p}) = \int \frac{d^{3}x}{(2\pi\hbar)^{3}}\nu({\mathbf x},{\mathbf p}).
\label{wig_mom}
\end{equation}

At $T=0$ the Wigner function becomes a Heaviside theta function and is given by
\begin{equation}
\nu({\mathbf x},{\mathbf p}) = \Theta[E_{F}-H({\mathbf x},{\mathbf p})].
\label{wig_func_zerot}
\end{equation}

In this limit, both the particle density and the momentum distribution assume a well-defined shape. The particle density is given for a general velocity independent potential $V({\bf x})$ by
\begin{equation}
n({\bf x})=\dfrac{\sqrt{2}\left[E_{F}-V({\bf x})\right]^{\frac{3}{2}}}{3\pi^{2}a_{z}^{3}\epsilon_{z}^{\frac{3}{2}}}\Theta\left[E_{F}-V({\bf x})\right].\label{eq:config-dist-1}
\end{equation}
From \eqref{eq:config-dist-1}, we see that a given rotation $\Omega_{\rm{DO}}$, at which the hole in the trap center appears, the system looks like a `donut', so to speak. The value of $\Omega_{\rm{DO}}$ can, therefore, be determined from \eqref{eq:EfDonut} by setting $E_F=0$ and yields
\begin{equation}
 {\Omega_{\rm{DO}}}={\omega_\perp}\sqrt{1 + \dfrac{(60\kappa^3N)^{\frac{1}{5}}}{\lambda^2}}.
\label{eq:omega_donut}
\end{equation}
In an effectively two-dimensional (uniform over an axial length $Z$), interacting BEC trapped by the radial components of the anharmonic potential \eqref{eq:pot-2}, rotation can lead as well to a hole in the center of the trap \cite{fetter:013605}, where, in the Thomas-Fermi approximation, a similar analytic expression can be written down:
\begin{equation}
 {\Omega_{\rm{h}}} = {\omega_\perp}\sqrt{1 + \dfrac{(12\kappa^2Na/Z)^{\frac{1}{3}}}{\lambda^2}},
\label{eq:omega_donut_BEC}
\end{equation}
where $a$ denotes the s-wave scattering length. In the case of non-interacting bosons, $\Omega_{\rm{h}}$ coincides with $\omega_\perp$, but this is not true for fermions, where an effective interaction arises from the Fermi statistics and $\Omega_{\rm{DO}}$ does not equal $\omega_{\perp}$ even in the absence of interactions. A discussion of this Pauli ``pseudopotential'' in the case of harmonically trapped Fermi gases is provided by Ref.~\cite{PhysRevA.55.4346}.

The value of $\Omega_{\rm{DO}}$ could also be obtained from the length scales $R^{\rm TF}_{i}$ set up by the Fermi energy, describing the extent of the gas in space,
\begin{equation}
\left.\left[V({\bf x})-E_{F}\right]\right|_{x_{i}=R_{i}^{\rm TF}}=0,
\label{eq:fermiR-1}
\end{equation}
where $i,\, j$ range over $x,\, y,\, z$. These length scales describe the maximum
extent of the coordinate space distributions at $T=0$ and are called the Thomas-Fermi
radii $R_{i}^{\rm TF}$. For the anharmonic potential \eqref{eq:pot-2}, they are given
by
\begin{eqnarray}
R_{\perp}^{\rm TF} & = & \dfrac{a_{z}}{\kappa}\left(-\lambda^{2}\eta\pm\sqrt{\lambda^{4}\eta^{2}+\dfrac{4\kappa E_{F}}{\epsilon_{z}}}\right)^{\frac{1}{2}},\nonumber\\
R_{z}^{\rm TF} & = & \sqrt{\dfrac{1}{\epsilon_{z}}\left[E_{F}+\theta(-\eta)\Delta E\right]},
\label{eq:R-TF}
\end{eqnarray}
where the positive root $R_{\perp}^{\rm TF}$ gives the outer radius of the distribution and the negative root has a physical meaning only in the `donut' regime, where it gives the inner radius. We note that for supercritical rotations ($E_{F}<0$), the density distribution has the shape of a shell, of which we then designate the outermost limit in the $z=0$-plane as the Thomas-Fermi radius $R_{\perp}^{\rm TF}$ and the largest extention in the $z$-direction as $R_{i}^{\rm TF}$. In the harmonic limit, the Thomas-Fermi radii are
\begin{equation}
R_{\perp}^{\rm TF,HO}=\dfrac{\left(48N\right)^{\frac{1}{6}}a_{z}}{\left(\lambda^{2}\eta\right)^{\frac{1}{3}}};\;\;\;\; R_{z}^{\rm TF,HO}=\left(48N\right)^{\frac{1}{6}}\left(\lambda^{2}\eta\right)^{\frac{1}{6}}.\label{eq:R-TF-HO}
\end{equation}

\begin{figure}
\begin{centering}
\includegraphics[width=7.9cm]{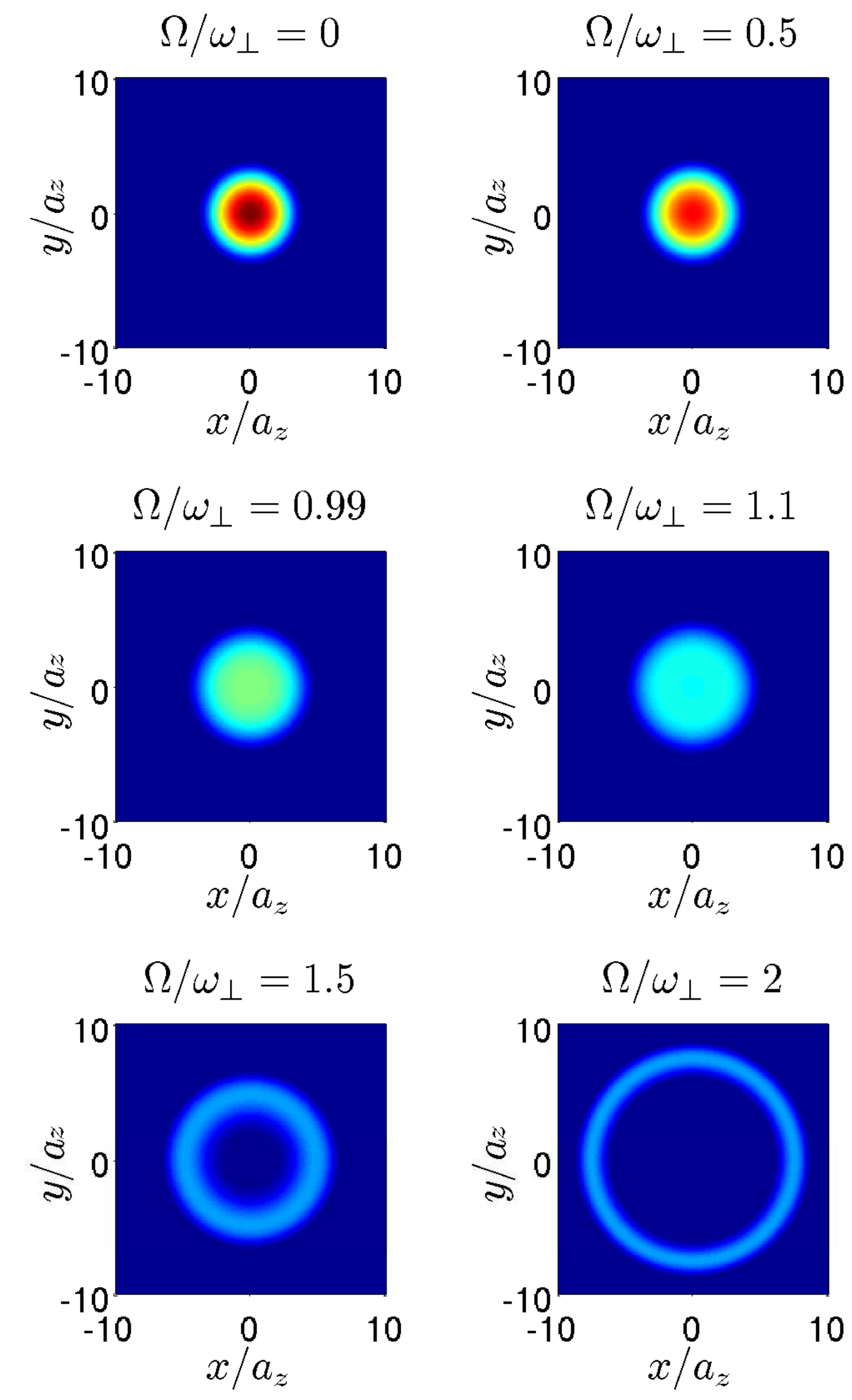}
\par\end{centering}
\caption{(Color Online) The $T=0$ spatial density distribution \eqref{eq:config-dist-2}
for the parameters of the Paris experiment at increasing rotations.
The anharmonic term prevents the formation of a `pancake' at near-critical
rotations and causes the gas to take on a `donut' shape for large
supercritical rotations.\label{fig:6}}
\end{figure}

In particular, we are interested in the optical density which can be directly measured by absorption images. Performing an integration over the axial $z$-direction, we obtain
\begin{equation}
n({ r_{\perp}})=\int n({r_{\perp}},\, z)dz=\dfrac{\left[E_{F}-V(r_{\perp},\, z=0)\right]^{2}}{4\pi a_{z}^{2}\epsilon_{z}^{2}}.\label{eq:config-dist-2}
\end{equation}
The rotation dependence of the $T=0$ column density \eqref{eq:config-dist-2}
is shown in Fig.~\ref{fig:6} for the anharmonic potential
\eqref{eq:pot-2}. Just above the critical rotation the distribution
becomes nearly homogeneous, and at high supercritical rotations the
gas eventually enters the `donut' regime, where the gas is entirely confined
to the dip in the trap. In contrast, without the quartic trapping
the gas spreads out to become an infinite pancake as the rotation
approaches the critical rotation, as we can see from the divergence
of the harmonic oscillator Thomas-Fermi radius \eqref{eq:R-TF-HO} at
the critical rotation.

\subsection{Momentum Distribution}
\label{sec:4D}
The Fermi energy also sets a momentum scale through the Fermi momentum $p_{F}=\sqrt{2M\left[E_{F}+\theta(-\eta)\Delta E\right]}$. It represents the maximum extent of the momentum distribution at $T=0$. For the anharmonic trap we have ${a_{z}}p_{F}={\hbar} \sqrt{\left[{E_{F}+\theta(-\eta)\Delta E}\right]/{\epsilon_{z}}}$.

A simple dependence on the particle number emerges again in the harmonic
limit
\begin{equation}
p_{F}^{\rm HO}=\dfrac{\hbar\left(48N\right)^{\frac{1}{6}}(\lambda^{2}\eta)^{\frac{1}{6}}}{a_{z}}.
\end{equation}

At $T=0$ the momentum space distribution \eqref{wig_mom} for the anharmonic potential
\eqref{eq:pot-2} reads
\begin{equation}
n({\bf p})=\dfrac{\sqrt{2}a_{z}^{3}}{16\pi^{2}\hbar^{3}}\begin{cases}
(q+1)\tan^{-1}(\sqrt{q})-\sqrt{q}; & \eta\geq0\\
(q+1)\left[\pi-\tan^{-1}(\sqrt{q})\right]+\sqrt{q}; & q\geq0\\
(q+1)\pi; & q\leq0\end{cases},\nonumber\\
\label{eq:mom-distr}
\end{equation}
with the dimensionless variable
\begin{equation}
q = \dfrac{E_{F}}{\Delta E}\left[1-\dfrac{{\mathbf p}^{2}}{2ME_{F}}\right].
\label{eq:xdef}
\end{equation}
We see that, in contrast to BEC, where the momentum distribution of the condensed atoms inherits the ani\-so\-tro\-py of the trap, the fermionic momentum distribution of a non-interacting gas remains isotropic despite the trap anharmonicity. This feature is drastically affected by presence of the anisotropic dipole-dipole interaction, as then the Fock term of the two-body mean-field energy induces a characteristic deformation of the momentum distribution \cite{miyakawa:061603}. Note that we have only considered the co-rotating frame; as we will see, the rotation leads to an anisotropy in the momentum distribution in the lab frame.

\subsection{Free Expansion}
\label{sec:4E}
Measurements in cold gases are usually carried out by turning off the trap and shining a resonant laser on the sample after some time of free expansion. These are called time-of-flight measurements (TOF).

So far we have treated the problem in the corotating frame. In considering TOF measurements, it is necessary to consider the laboratory frame, where the gas expands ballistically.  The transformation of the momenta from the corotating frame to the lab frame yields
\begin{equation}
 {\mathbf p}_{\rm lab} = \mathbf{p} + M\Omega r {{\bf \hat{\phi}}},
\end{equation}
where $\bf\hat{\phi} $ is the tangential unit vector.
Since the trap is radially symmetric, the coordinate transformation of the position does not affect the 
distribution. We have then
\begin{equation}
n_{\rm lab}({\mathbf x}_{\rm lab}, {\mathbf p}_{\rm lab}) = n(\mathbf{x}, \mathbf{p}) = {n_{\rm lab}}(\mathbf{x}_{\rm lab}, {\mathbf p}_{\rm lab} - M\Omega r{{\hat{\phi}}}).
\end{equation}
Thus although the momentum distribution does not inherit the anisotropy of the trap, the trap rotation leads nonetheless to an anisotropic momentum distribution.

To study the expansion in detail we apply the ballistic substitution
\begin{equation}
({\mathbf x},{\mathbf p}) \rightarrow ({\mathbf x}-\frac{{\mathbf p}t}{M},{\mathbf p}),
\end{equation}
which yields a relation between the lab frame distribution and the corotating distribution:
\begin{equation}
n_{\rm lab}({\mathbf x}_{\rm lab},{\mathbf p}_{\rm lab}, t) = n\left[{\mathbf x}_{\rm lab}-\frac{{\mathbf p}_{\rm lab}t}{M},{\bf p}_{\rm lab}-M\Omega r_{\perp}^{(0)}\hat{\phi}_0\right],
\end{equation}
where $r_{\perp}^{(0)}$ and $\hat{\phi}_0$ denote respectively the radius and tangential vector at the position ${\bf x}_0 = {\bf x}_{\rm lab}-\frac{{\bf p}_{\rm lab}t}{M}$, and can be expressed as 
\begin{equation}
r_{\perp}^{(0)}\hat{\phi}_0 = {\bf\hat{z}}\times{\bf x}_0 = {\bf \hat{z}}\times\left({\bf x}_{\rm lab}-\frac{{\bf p}_{\rm lab}t}{M}\right),
\end{equation}
with $\bf \hat{z}$ the unit vector in the axial direction.

Of particular experimental interest is the density distribution
at a time $t$ after the trap is turned off,
\begin{eqnarray}
n_{\rm lab}({\bf x},\, t) & = & \int \frac{d^{3}p_{\rm lab}}{(2\pi\hbar)^{3}}\, \Theta\left[E_{F} -\frac{ ({\mathbf p}_{\rm lab}-M\Omega r_{\perp}^{(0)}\hat{\phi}_0)^{2}}{2M}\right.\nonumber\\
& &\left. - V\left({\bf x}-\dfrac{{\mathbf p}_{\rm lab}t}{M}\right)\right].\label{eq:ballistic}
\end{eqnarray}
For the case of harmonic trapping and zero rotation, the kinetic term can be combined
with the harmonic term and the integral over momentum space in \eqref{eq:ballistic}
takes nearly the same form as the integration over momentum space
in \eqref{eq:g-1b}. Unfortunately, in the case of the anharmonic
potential \eqref{eq:pot-2} with rotation, the integration can not be cast into
a simple form and does not provide an insightful analytic expression.

A measure of the spatial density is given by the aspect ratio
\begin{equation}
\varepsilon(t)=\dfrac{W_{z}(t)}{W_{\perp}(t)},\label{eq:aspect ratio}
\end{equation}
with the respective widths
\begin{equation}
W_{i}(t)=\left[\int d^{3}xx_{i}^{2}n_{\rm lab}({\bf x},\, t)\right]^{\frac{1}{2}}.\label{eq:Wi}
\end{equation}
For harmonically confined Fermi gases with zero rotation, it is known that anisotropic samples become asymptotically isotropic for large expansion time \cite{PhysRevLett.89.250402}, since their widths scale in time as $\omega_{i}^{-1}(1+\omega_{i}^{2}{t}^{2})^{\frac{1}{2}}$ and the aspect ratio as
\begin{equation}
\varepsilon(t) = \sqrt{\frac{\omega_{\perp}^{2}} {\omega_{z}^{2} }} \sqrt{\frac{ 1 + \omega_{z}^{2} {t}^{2}} { 1 + \omega_{\perp}^{2} {t}^{2}}}.
\end{equation}

\begin{figure}
\begin{centering}
\includegraphics[width=8cm]{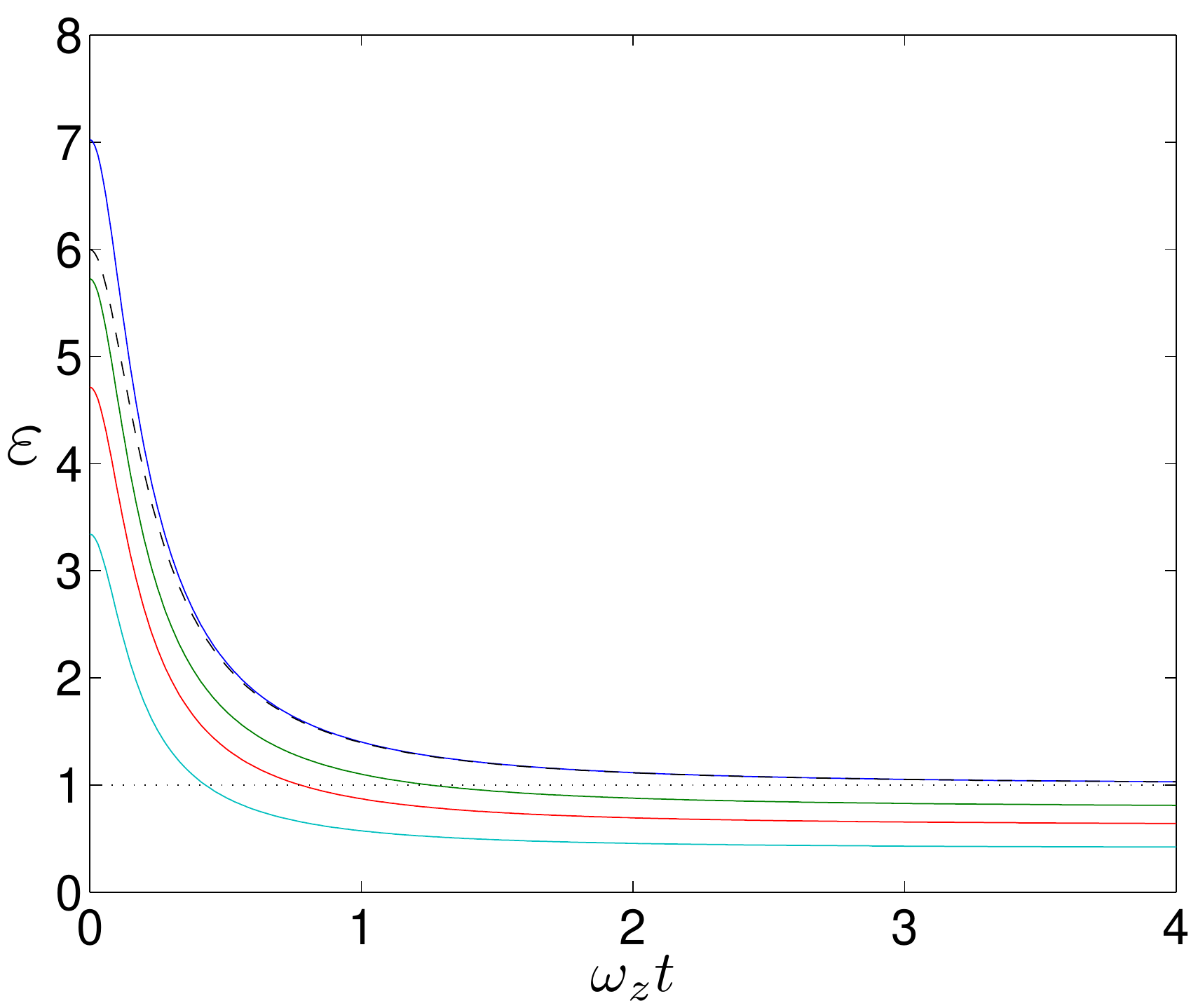}
\par\end{centering}
\caption{(Color online) The aspect ratio \eqref{eq:aspect ratio} of the spatial density for the parameters of the Paris experiment at different rotation frequencies as a function of expansion time. The dashed curve is for the harmonic trap at zero rotation, and the solid curves are for the anharmonic trap at rotations of $\Omega/\omega_\perp = 0,\,0.75,\,0.99,\,1.25$ from top to bottom.\label{fig:7}}
\begin{centering}

\par\end{centering}
\end{figure}

In the case of a rotating gas with additional quartic trap, it is interesting to see how the rotation affects the asymptotic value of the aspect ratio, which reflects the lab frame momentum distribution. In Fig.~\ref{fig:7} the aspect ratio \eqref{eq:aspect ratio} has been evaluated numerically for the anharmonic potential. Comparing to the harmonic trap at zero rotation, we see that the anisotropy of the momentum distribution increases with the rotation frequency for the anharmonic trap. It is worth mentioning that the value of $\varepsilon(t)$ for large asymptotically large $t$ depends on the rotation, so that the aspect ratio could be used to measure the actual rotation frequency in an experiment.

\section{Temperature Effects}
\label{sec:5}
After having studied the effects of rotation on the ground state of an anharmonically trapped polarized Fermi gas, we turn our attention to the thermodynamic properties of such a system. The standard thermodynamical recipe is specialized to the case of the anharmonic potential and the consequences for different rotation frequencies are explored in both low- and high-temperature regimes.

\subsection{Standard Grand-Canonical Approach}
\label{sec:5A}
In this section we describe the thermodynamical properties of the system within the framework of the grand-canonical ensemble, where the most fundamental quantity is the grand-canonical partition function. For particles obeying Fermi-Dirac statistics it is given by
\begin{equation}
{\cal Z}_{G}=\prod_{n}\left[1+e^{-\beta(E_{n}-\mu)}\right],
\label{eq:Zg}
\end{equation}
where $E_{n}$, as before, is the eigenvalue corresponding to the eigenstate $\ket{n}$.

The grand-canonical free energy is obtained from the grand-canonical
partition function through
\begin{equation}
{\cal F}_{G}=-\dfrac{1}{\beta}\ln{\cal Z}_{G}.\label{eq:Fg-2}
\end{equation}
Inserting \eqref{eq:Zg} in \eqref{eq:Fg-2} and making use of the semiclassical approximation, i.e., writing the discrete sum as an integral over the density of states \eqref{eq:g-1a}, we obtain
\begin{eqnarray}
{\cal F}_{G} & = & {-\dfrac{1}{\beta}} \int_{-\infty}^{\infty}g(E)\ln\left[1+e^{-\beta(E-\mu)}\right]dE.\label{eq:Fg-3}\end{eqnarray}
Performing a partial integration of \eqref{eq:Fg-3}, one sees that the integral for the grand-canonical free energy contains
the Fermi distribution:
\begin{equation}
{\cal F}_{G}=-{\dfrac{1}{\beta}}\int_{-\infty}^{\infty}G(E)\frac{dE}{e^{\beta(E-\mu)}+1},
\label{eq:Fg-4}
\end{equation}
where $G(E)$ denotes the number of single-particle states with energy less than $E$ 
\begin{equation}
G(E)=\int_{-\infty}^{E}g(E')dE'.
\end{equation}

In view of using the semiclassical approximation for the density of
states in \eqref{eq:g-2}, we define the incomplete Fermi integral
\begin{equation}
\zeta_{\nu}^{+,{\rm inc}}(z,x_{0})=\frac{1}{\Gamma(\nu)}\int_{x_0}^{\infty}\frac{x^{\nu-1}}{1+e^{x}z^{-1}}dx,
\label{eq:zetap-inc}
\end{equation}
which is called incomplete due to the fact that its integration range is smaller than the one of the original Fermi integral, defined by
\begin{equation}
\zeta_{\nu}^{+}(z)=\frac{1}{\Gamma(\nu)}\int_{0}^{\infty}\frac{x^{\nu-1}}{1+e^{x}z^{-1}}dx=\zeta_{\nu}^{+,{\rm inc}}(z,0).\label{eq:zetap}
\end{equation}
For convenience we introduce the further shorthand
\begin{eqnarray}
\zeta_{\nu}^{+,{\rm inc}} & = & \zeta_{\nu}^{+,{\rm inc}}\left(e^{\beta(\mu+\Delta E) },\beta\Delta E\right),\nonumber\\
\zeta_{\nu}^{+}& = &\zeta_{\nu}^{+}\left(e^{\beta(\mu+\Delta E) }\right).
\end{eqnarray}

With \eqref{eq:g-2} and \eqref{eq:Fg-4}, we write the grand-canonical
free energy of a system contained in the anharmonic potential \eqref{eq:pot-2}
in terms of the following Fermi integrals:
\begin{eqnarray}
{\cal F}_{G} & = & -\dfrac{\sqrt{\pi}\left[2\Theta(-\eta) \zeta_{\frac{7}{2}}^{+} + \sigma(\eta) \zeta_{\frac{7}{2}}^{+,{\rm inc}} \right]} {2\beta^{\frac{7}{2}} \epsilon_{z}^{\frac{5}{2}} \kappa^{\frac{1}{2}}}  + \dfrac{\lambda^{2}\eta\zeta_{3}^{+,{\rm inc}}} {2\beta^{3} \epsilon_{z}^{2}\kappa} \nonumber \\
&  &  - \dfrac{\lambda^{6} \eta^{3}\zeta_{2}^{+,{\rm inc}}} {24\beta^{2} \epsilon_{z} \kappa^{2}}  + \dfrac{\lambda^{10} \eta^{5}\zeta_{1}^{+,{\rm inc}}} {320\beta\kappa^{3}} ,
\label{eq:Fg-5}
\end{eqnarray}
where $\sigma(\eta)$ is the sign of $\eta$, defined as
\begin{equation}
\sigma(\eta) =
\begin{cases}
+1; & \eta \ge 0\\
-1; & \eta < 0\end{cases}.
\end{equation}

To obtain the other thermodynamical quantities, we express the grand-canonical free energy in terms of
a Legendre transformation 
\begin{equation}
{\mathcal F}_{G}=U-TS-\mu N,\label{eq:Fg-1}
\end{equation}
with the differential
\begin{equation}
d{\cal F}_{G} = dU-SdT - Nd\mu.\label{eq:Fg-1_2}
\end{equation}
Therefore, the particle number $N$ and entropy $S$ can thus be determined
from the grand-canonical free energy according to
\begin{equation}
N=-\left.\dfrac{\partial {\mathcal F}_{G}}{\partial\mu}\right|_{T},\;\;\;\; S=-\left.\dfrac{\partial {\mathcal F}_{G}}{\partial T}\right|_{\mu}.
\end{equation}
For the particle number we obtain
\begin{eqnarray}
N & = & \dfrac{\sqrt{\pi}\left[2\Theta(-\eta) \zeta_{\frac{5}{2}}^{+}+\sigma(\eta)\zeta_{\frac{5}{2}}^{+,{\rm inc}}\right]} {2\beta^{\frac{5}{2}} \epsilon_{z}^{\frac{5}{2}} \kappa^{\frac{1}{2}}}-\dfrac{\lambda^{2}\eta\zeta_{2}^{+,{\rm inc}}}{2\beta^{2}\epsilon_{z}^{2}\kappa}\nonumber\\
& &  + \dfrac{\lambda^{6}\eta^{3}\zeta_{1}^{+,{\rm inc}}}{24\beta\epsilon_{z}\kappa^{2}},
\label{eq:N}
\end{eqnarray}
and for the entropy we get
\begin{eqnarray}
{TS} & = & -{\mathcal F}_{G}-(\mu+\Delta E) N+ \dfrac{5\sqrt{\pi}\Theta(-\eta) \zeta_{\frac{7}{2}}^{+} } {2\beta^{\frac{7}{2}} \epsilon_{z}^{\frac{5}{2}} \kappa^{\frac{1}{2}}}  \nonumber\\
&  & + \dfrac{5\sqrt{\pi} \sigma(\eta) \zeta_{\frac{7}{2}}^{+,{\rm inc}}} {4\beta^{\frac{7}{2}} \epsilon_{z}^{\frac{5}{2}} \kappa^{\frac{1}{2}}} - \dfrac{\lambda^{2}\eta \zeta_{3}^{+,{\rm inc}}} {\beta^{3} \epsilon_{z}^{2}\kappa} + \dfrac{\lambda^{6} \eta^{3} \zeta_{2}^{+,{\rm inc}}} {6\beta^{2}\epsilon_{z}}. \label{eq:S}
\end{eqnarray}

The internal energy can then be obtained from \eqref{eq:Fg-1} yielding
\begin{eqnarray}
U & = & \dfrac{5\sqrt{\pi}\left[ 2\Theta(-\eta) \zeta_{\frac{7}{2}}^{+} + \sigma(\eta)\zeta_{\frac{7}{2}}^{+,{\rm inc}}\right]} {4\beta^{\frac{7}{2}} \epsilon_{z}^{\frac{5}{2}}\kappa^{\frac{1}{2}}}-\dfrac{\lambda^{2}\eta\zeta_{3}^{+,{\rm inc}}}{\beta^{3}\epsilon_{z}^{2}\kappa} \nonumber\\
& &  + \dfrac{\lambda^{6}\eta^{3}\zeta_{2}^{+,{\rm inc}}}{6\beta^{2}\epsilon_{z}}-N\Delta E.
\label{eq:U}
\end{eqnarray}

Notice that, in the pure quartic limit $(\eta=0)$, a simple relation between the free and the internal energies arises
\begin{equation}
U = -\frac{5}{2}{\mathcal F}_{G}.
\end{equation}

The heat capacity of the system at constant particle number and trap parameters is given by
\begin{equation}
C_{}=\left.\dfrac{\partial U}{\partial T}\right|_{N}.
\end{equation}
To evaluate this derivative, we must, as usual, obtain the temperature derivative of the chemical potential from the 
With \eqref{eq:U} we then obtain

{\tiny
\begin{eqnarray}
\frac{C}{k_{B}} & = & \dfrac{35\sqrt{\pi}\left[2\Theta(-\eta) \zeta_{\frac{7}{2}}^{+} + \sigma(\eta)\zeta_{\frac{7}{2}}^{+,{\rm inc}} \right]}{8\beta^{\frac{5}{2}} \epsilon_{z}^{\frac{5}{2}} \kappa^{\frac{1}{2}}} - \dfrac{3\lambda^{2}\eta\zeta_{3}^{+,{\rm inc}}} {\beta^{2}\epsilon_{z}^{2} \kappa} + \dfrac{\lambda^{6} \eta^{3}\zeta_{2}^{+,{\rm inc}}} {12\beta \epsilon_{z}\kappa^{2}} \nonumber\\
&  & -\dfrac{\left\{\dfrac{5\sqrt{\pi}\left[ 2\Theta(-\eta) \zeta_{\frac{5}{2}}^{+} + \sigma(\eta) \zeta_{\frac{5}{2}}^{+,{\rm inc}} \right]}{\beta^{\frac{5}{2}}\epsilon_{z}^{\frac{5}{2}}\kappa^{\frac{1}{2}}} - \dfrac{4\lambda^{2}\eta\zeta_{2}^{+,{\rm inc}}} {\beta^{2} \epsilon_{z}^{2}\kappa} + \dfrac{\lambda^{6} \eta^{3}\zeta_{1}^{+,{\rm inc}}} {6\beta \epsilon_{z} \kappa^{2}} \right\}^{2}} {\dfrac{ 8\sqrt{\pi}\left[2\Theta(-\eta) \zeta_{\frac{3}{2}}^{+} + \sigma(\eta)\zeta_{\frac{3}{2}}^{+,{\rm inc}} \right]} {\beta^{\frac{5}{2}} \epsilon_{z}^{\frac{5}{2}} \kappa^{\frac{1}{2}}}  - \dfrac{8\lambda^{2} \eta\zeta_{1}^{+,{\rm inc}}}{\beta^2\epsilon_{z}^{2}\kappa}}\nonumber\\
\label{eq:Cv-sub}
\end{eqnarray}}

\begin{figure}
\begin{centering}
\includegraphics[width=8cm]{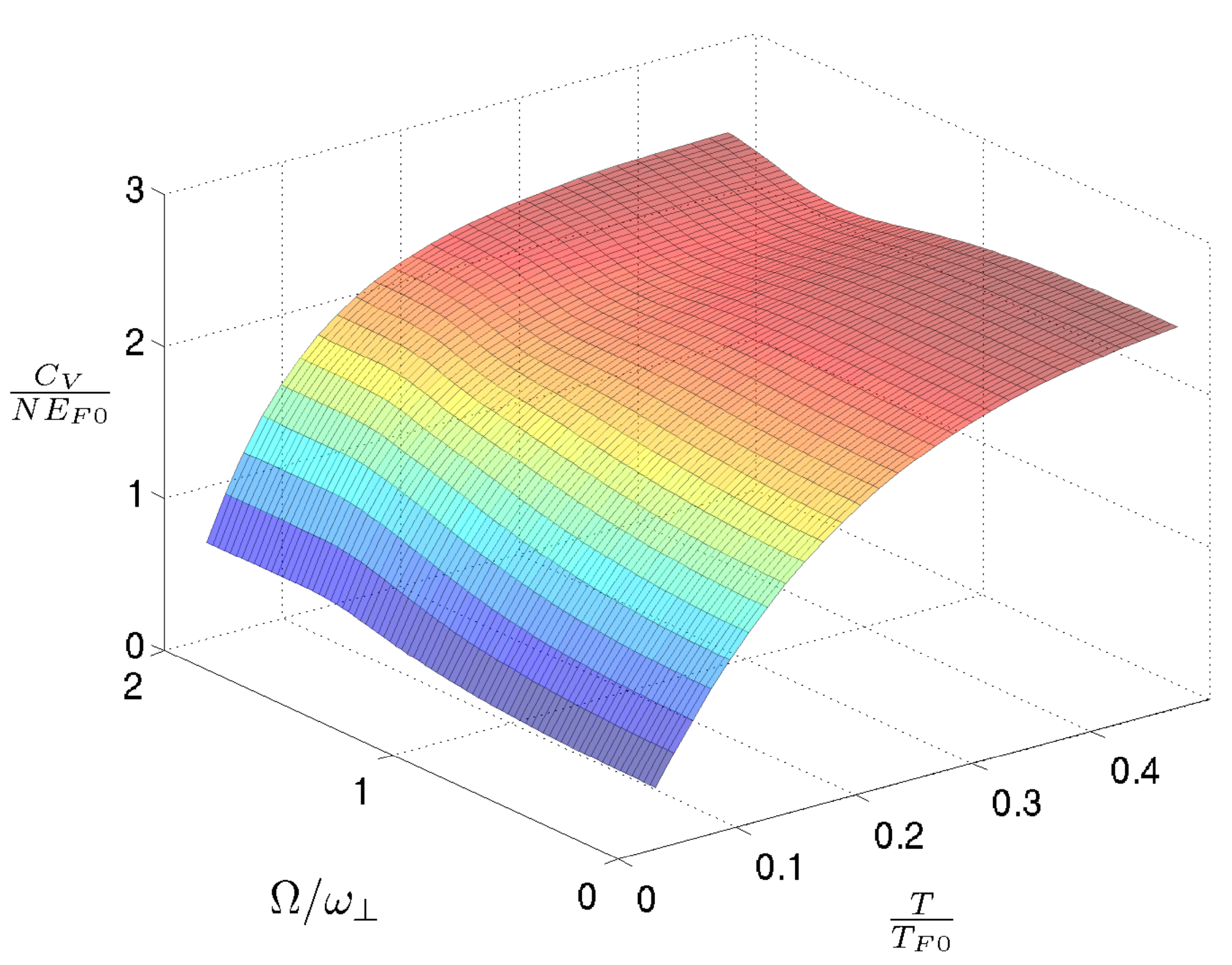}
\par\end{centering}
\caption{(Color Online) Heat capacity per particle at constant particle number and trap parameters for varying temperatures and rotation speeds from \eqref{eq:Cv-sub}. The energy scale is $E_{f0}$, the Fermi energy of the system at zero rotation ($\Omega=0)$. The
temperature scale is the Fermi temperature, $T_{f0}=E_{f0}/k_{B}$.
The angular frequency scale is the critical frequency $\omega_{\perp}$.\label{fig:8}}
\end{figure}

Since this analytic expression is not particularly illuminating, we analyze the behavior of \eqref{eq:Cv-sub} for a range of temperatures from Fig.~\ref{fig:8}. At low temperatures the heat capacity is nearly linear in temperature, and at high temperatures the heat capacity approaches $C_{}={5}Nk_{B}/{2}$ for a fixed rotation. At high temperatures a dip appears with a minimum occurring at a rotation that increases with the temperature. This interesting case and the general high-temperature behavior of the heat capacity will be discussed in detail below.

\subsection{Low-Temperature Approximation}
\label{sec:5B}
In the low-temperature regime, we can understand temperature effects by using the Sommerfeld expansion as sketched in Ref. \cite{ashcroft-ssp}. To this end we consider the incomplete Fermi function with $X=\mu+\Delta E$
\begin{equation}
\zeta_{\nu}^{+,{\rm inc}}(e^{\beta X},\beta \Delta E)=\frac{\beta^{\nu}}{\Gamma(\nu)}\int_{\Delta E}^{\infty}E^{\nu-1}f(E,\,\beta,\,X )dE,\label{eq:fermi-som-1}
\end{equation}
where $f(E,\,\beta,\,X)$ is the Fermi distribution \eqref{eq:fermi-distribution}.
A partial integration yields
\begin{eqnarray}
\zeta_{\nu}^{+,{\rm inc}}(e^{\beta X },\beta \Delta E) & = & \dfrac{\beta^{\nu}}{\Gamma(\nu)} \left[-\dfrac{1}{\nu}f(\Delta E,\,\beta,\,X) \Delta E^{\nu}\right.\nonumber\\
& & \left.-\int_{\Delta E}^{\infty}E^{\nu}\,\dfrac{df(E,\,\beta, X)}{dE}dE\right],\label{eq:fermi-som-1a}
\end{eqnarray}
where the derivative of the Fermi distribution is given by
\begin{equation}
\dfrac{df(E,\,\beta,\,X)}{dE}=-\dfrac{\beta}{\left[e^{\beta(E-X )}+1\right]\left[e^{-\beta(E-X)} + 1 \right] }.\label{eq:df}
\end{equation}

Consider the case that $\left|E-X \right|\gg k_{B}T$. We see that the derivative \eqref{eq:df} becomes very small. If $X > 0$, then the lower limit of integration
in \eqref{eq:fermi-som-1a} can be extended to $-\infty$ and $f(\Delta E,\,\beta,\,X)\approx1$. If $X < 0$
then the integrand vanishes and $f(\Delta E,\,\beta,\,X)\approx0$. We now expand
$E^{\nu}$ in a Taylor series about $E=X $,
\begin{eqnarray}
E^{\nu} & = & \sum_{n=0}^{\infty}\left[\dfrac{\Gamma(\nu+1)}{\Gamma(\nu+1-n)}X^{\nu-n}\right]\frac{(E-X )^{n}}{\Gamma(n+1)}\nonumber\\
\label{eq:fermi-taylor}
\end{eqnarray}
and integrate \eqref{eq:fermi-som-1} with \eqref{eq:fermi-taylor}
to obtain
\begin{equation}
\frac{\zeta_{\nu}^{+,{\rm inc}}(e^{\beta X},\beta \Delta E)}{\beta^{\nu}}  \approx 
\dfrac{ X^{\nu}\! -\!  \Delta E^{\nu}} {\Gamma(\nu+1)} + X ^{\nu}\sum_{n=1}^{\infty} \dfrac{a_{n}\left(\beta X \right)^{-2n}} {\Gamma(\nu+1-2n)}
\label{eq:fermi-inc-low}
\end{equation}
It is worth remarking that the Sommerfeld expansion (\ref{eq:fermi-inc-low}) is valide for $X>0$ and vanishes for $X<0$. Furthermore, the $a_{n}$ are constants given in terms of the Riemann zeta function
$\zeta(n)$ as 
\begin{equation}
a_{n}=\left[2-\dfrac{1}{2^{2(n-1)}}\right]\zeta(2n).
\end{equation}
For the case of the complete Fermi integral, Eq.~\eqref{eq:fermi-inc-low}
reduces to the usual Sommerfeld expansion given by
\begin{equation}
\zeta_{\nu}^{+}(e^{\beta X})  \approx  \dfrac{\left( \beta X \right)^{\nu}} {\Gamma(\nu+1)}+ \left( \beta X \right)^{\nu} \sum_{n=1}^{\infty} \dfrac{a_{n}\left(\beta X \right)^{-2n}} {\Gamma(\nu+1-2n)},
\label{eq:fermi-low}
\end{equation}
for $X>0$.

Although the series in \eqref{eq:fermi-inc-low} and \eqref{eq:fermi-low}
terminate for $\nu=0,1,2...$, the expansion is not exact since the
limits of integration were extended in \eqref{eq:fermi-som-1}.

\begin{figure}
\begin{centering}
\includegraphics[width=8cm]{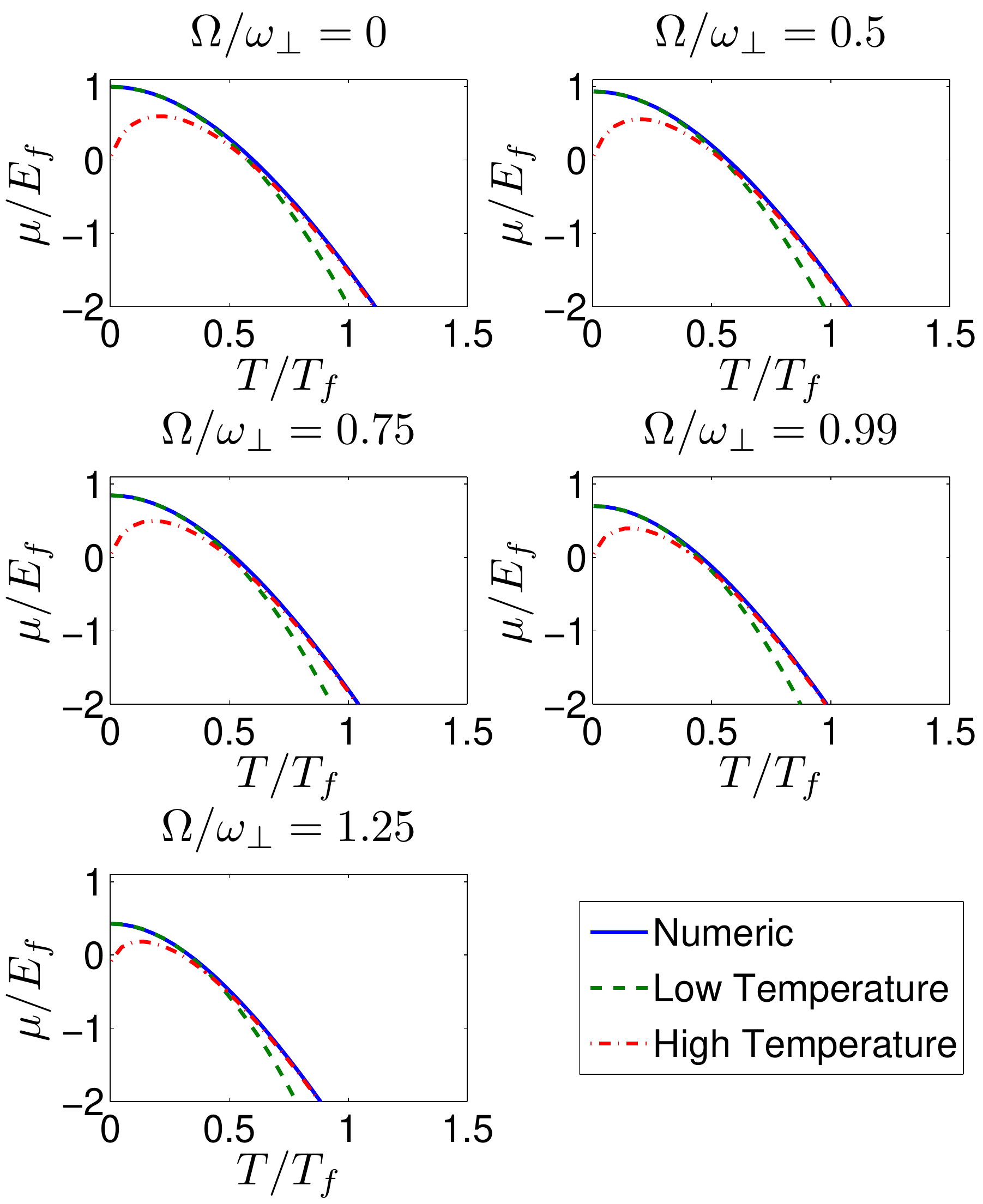}
\par\end{centering}
\caption{(Color online) Evaluation of chemical potential from particle number equation. A
numeric inversion of \eqref{eq:N} is shown for the general case. For the low
temperature expansion \eqref{eq:muLow} is shown, and for the high temperature expansion
\eqref{eq:HighT-N} has been solved analytically for $\mu$.\label{fig:9}}
\end{figure}

It is useful to obtain an explicit equation for the temperature dependence
of the chemical potential using the Sommerfeld expansion \eqref{eq:fermi-inc-low} and the particle number equation \eqref{eq:N}. To this end, we  substitute $\mu=E_{F}+\delta$ and keep terms linear in $\delta$ as well as quadratic in temperature to obtain up to third order contributions
\begin{equation}
\mu=E_{F}-\dfrac{\pi^{2}}{36g_{0}(E_{F})}\left[\dfrac{6\left(E_{F}+\Delta E \right)^{\frac{1}{2}}} {\epsilon_{z}^{\frac{5}{2}}\kappa^{\frac{1}{2}}} - \dfrac{3\lambda^{2}\eta} {\epsilon_{z}^{2}\kappa} \right] (k_{B}T)^{2},
\label{eq:muLow}
\end{equation}
where $g_{0}(E_{F})$ is the density of states \eqref{eq:g-2} evaluated
at the Fermi energy. We see that for low temperatures the chemical
potential decreases with the square of the temperature. This result
is compared to the numerical inversion of the particle number equation
\eqref{eq:N} with both the exact evaluation of the Fermi integrals
and with the evaluation using terms quartic in temperature from the Sommerfeld
expansion in Fig.~\ref{fig:9}.

\subsection{High-Temperature Approximation\label{GrandCan_hight}}
\label{sec:5C}
In the high-temperature limit we expect quantum statistics for both bosons and fermions to approach classical statistics. Here we derive the high-temperature expansions of the thermodynamic properties of a Fermi gas in the anharmonic trap of the Paris experiment. We also consider in more detail the unique rotation dependence of the heat capacity in the high-temperature limit.

A high-temperature expansion of the incomplete Fermi integrals is obtained as follows. We rewrite \eqref{eq:zetap-inc} as 
\begin{equation}
\zeta_{\nu}^{+,{\rm inc}} (z,\beta\Delta E) = \frac{1}{\Gamma(\nu)} \int_{\beta\Delta E}^{\infty} x^{\nu-1} \frac{e^{-x}z}{1+e^{-x} z}dx,
\end{equation}
where $z=e^{\beta(\mu + \Delta E)}$.
As seen in Fig.~\ref{fig:9}, if the temperature is high enough, $e^{\beta\mu}<1$ and, thus, $e^{-\beta\Delta E}z<1$ so that we can use the binomial expansion for the factor $\left(1+e^{-x}z\right)^{-1}$. Then we obtain
\begin{equation}
\zeta_{\nu}^{+,{\rm inc}}(z,\beta\Delta E)=\sum_{j=1}^{\infty}Q(\nu,j\beta\Delta E)\dfrac{z^{j}(-1)^{j-1}}{j^{\nu}},
\label{eq:HighT-Inc}
\end{equation}
with the definition
\begin{equation}
Q(\nu,\beta\Delta E)=\frac{1}{\Gamma(\nu)}\int_{\beta\Delta E}^{\infty}x^{\nu-1}e^{-x}dx,
\end{equation}
where $\Gamma(x)$ is the Gamma function.

For $z\rightarrow0$ in the high-temperature limit, it is sufficient
to keep only the first term in the sum, yielding
\begin{eqnarray}
\lim_{T\rightarrow\infty}\zeta_{\nu}^{+,{\rm inc}} & = & \lim_{T\rightarrow\infty}\zeta_{\nu}^{+,{\rm inc}}(e^{\beta(\mu+\Delta E) },\beta\Delta E)\nonumber\\
&\approx& e^{\beta(\mu+\Delta E) }\, Q(\nu,\beta\Delta E).
\label{eq:f-exp1}
\end{eqnarray}
For positive integer $n$, $Q(n,\beta\Delta E)$ can be expanded exactly:
\begin{equation}
Q(n,\beta\Delta E)=e^{\beta\Delta E}\sum_{j=0}^{n-1}\frac{(\beta\Delta E)^{j}}{j!}.
\end{equation}

The high-temperature expansion is compared to the exact evaluation
of the Fermi integrals in Fig.~\ref{fig:9} in the context
of a numerical inversion of the particle number equation to solve
for the chemical potential. The approximation is in general very accurate
for temperatures larger than half the Fermi temperature.

From \eqref{eq:f-exp1} we can determine the high-temperature behavior of the general expressions for the thermodynamic properties of the system. For the particle number, we get
\begin{equation}
N  =  \dfrac{e^{\beta(\mu+\Delta E)}}{2(\epsilon_{z}\beta)^{\frac{5}{2}}\kappa^{\frac{1}{2}}} 
 \left[2 \Theta(-\eta) + \sigma(\eta)Q\left(\frac{5}{2},\beta{\Delta E}\right)\right]
+\cdots.
\label{eq:HighT-N}
\end{equation}
It should be remarked that (\ref{eq:HighT-N}) is to be understood as giving the temperature dependence of the chemical potential in the high-temperature regime, so that $\mu$ is changed to maintain a constant value of $N$ on the left-hand side of the equation. Note that the high-temperature chemical potential can be determined explicitly by taking the logarithm of \eqref{eq:HighT-N}.

Now we are in a position to write down the high-temperature versions of the internal energy \eqref{eq:U}
\begin{equation}
\frac{U}{Nk_{B}T}  =  \dfrac{5}{2}-\beta{\Delta E}+\dfrac{\sigma(\eta)\pi^{-1/2}(\beta\Delta E)^{\frac{1}{2}} e^{-\beta{\Delta E}}}{\left[2\Theta(-\eta)+\sigma(\eta)Q\left(\frac{5}{2},\beta{\Delta E}\right)\right]}+\cdots
\label{eq:HighT-U}
\end{equation}
and of the heat capacity \eqref{eq:Cv-sub}
\begin{eqnarray}
\frac{C_{}}{Nk_{B}} & = & \dfrac{5}{2} + \left[\dfrac{1}{2}+\beta{\Delta E}\right]\dfrac{\sigma(\eta)(\beta\Delta E)^{\frac{1}{2}}\pi^{-1/2} e^{-\beta{\Delta E}}}{\left[2\Theta(-\eta)+\sigma(\eta)Q\left(\frac{5}{2},\beta{\Delta E}\right)\right]}\nonumber\\& & -\dfrac{\beta{\Delta E}e^{-2\beta{\Delta E}}}{\pi\left[2\Theta(-\eta)+\sigma(\eta)Q\left(\frac{5}{2},\beta{\Delta E}\right)\right]^{2}} +\cdots\;.
\label{eq:HighT-Cv}
\end{eqnarray}

\begin{figure}
\begin{centering}
\includegraphics[width=8cm]{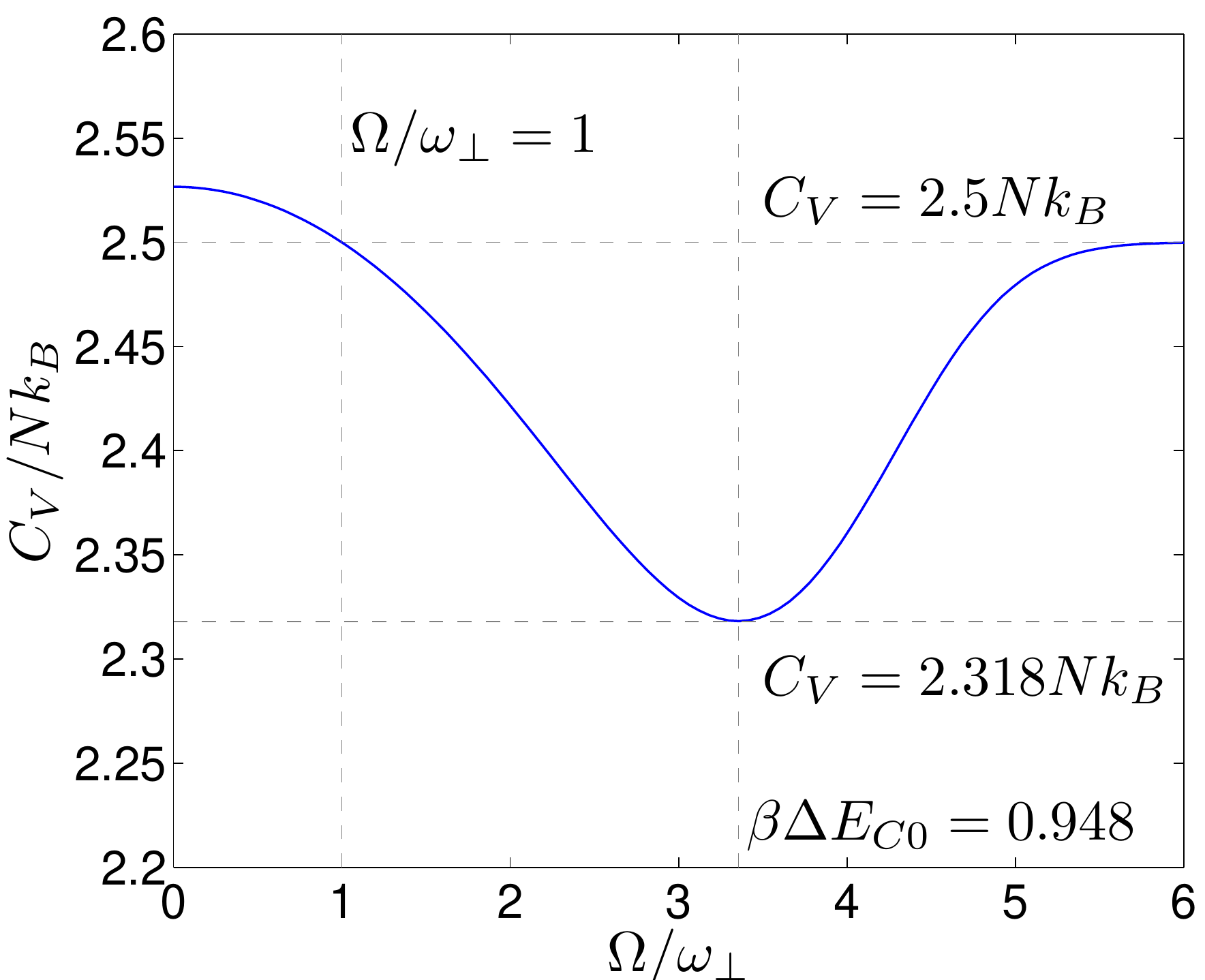}
\par\end{centering}
\caption{The high-temperature behavior of the heat capacity
as a function of rotation, evaluated using the parameters of the Table \ref{tab:1} and $T=1\,\mu{\rm K}$ with \eqref{eq:HighT-Cv}. From Eq.~(\ref{eq:Cv-min}), one obtains that the minimum of the heat capacity at $T=1\,\mu{\rm K}$ is located at $\Omega_{0}\approx3.43~\omega_{\perp}$.\label{fig:10}}
\end{figure}

The high-temperature behavior of the heat capacity \eqref{eq:HighT-Cv}
is shown in Fig.~\ref{fig:10}. For subcritical rotations we
have $C_{}>2.5~Nk_{B}$, for critical rotations $C_{}=2.5\, Nk_{B}$,
and for supercritical rotations $C_{}<2.5~Nk_{B}$. For a fixed
rotation, the heat capacity approaches in all cases $C_{}=2.5~Nk_{B}$
as $T\rightarrow\infty$. For a large constant temperature, the heat
capacity also approaches $C_{}=2.5~Nk_{B}$ in the high rotation
limit, $\Omega/\omega_{\perp}\rightarrow\infty$. 

The most interesting feature of the high-temperature heat capacity is the dip found in the supercritical region. Differentiation of \eqref{eq:HighT-Cv} with respect to $\Omega$ shows that the minimum heat capacity occurs at $\beta\Delta E_{C0}\approx 0.948$. Interestingly, the value of the minimum heat capacity, $C_{0} \approx 2.318 N k_{B}$, is independent of temperature and rotation. The relation between the temperature of the system and the rotation $\Omega_{0}$ corresponding to the minimum heat capacity can be written more clearly as 

\begin{equation}
{\Omega_0} = {\omega_\perp}\sqrt{1 + \left(\dfrac{3.792\kappa k_B T} {\epsilon_z\lambda^{4}}\right)^{\frac{1}{2}}}. 
\label{eq:Cv-min}
\end{equation}
The rotation $\Omega_{0}$ at which the minimum heat capacity is located increases monotonically with temperature and has the asymptotic form $\Omega_{0}\varpropto T^{1/4}$. Thus, although the minimum is present at any temperature, it occurs at increasingly high rotations so that the heat capacity flattens to $C_{}=2.5~Nk_{B}$ as $T\rightarrow\infty$ for any finite rotation. Furthermore, the similarity of \eqref{eq:Cv-min} to \eqref{eq:omega_donut} and \eqref{eq:omega_donut_BEC}, which give the critical rotation for the transition to the zero temperature donut regime, suggests that the minimum of the heat capacity is related to the transition of the gas into a configuration where the particles are mostly contained in the minimum of the supercritical potential. Recalling that $\Delta E$ is the potential height at the trap mininum, one should expect that the ``donut regime'' is realized when the thermal energy of the particles is of the order of $\Delta E$. Thus, setting $\Delta E \approx k_{B}T$ should provide an estimate of $\Omega_0$. Indeed, this gives
\begin{equation}
{\Omega_0} \approx {\omega_\perp}\sqrt{1 + \left(\dfrac{4\kappa k_B T} {\epsilon_z\lambda^{4}}\right)^{\frac{1}{2}}}, 
\label{eq:Cv-min-esti}
\end{equation}
which is not bad at all.

\subsection{Virial-Theorem Approach}
\label{sec:5D}
At high enough temperatures, the quantum mechanical effects may be neglected. Thus, the high-temperature properties of systems in an anharmonic trap can also be explained in the context of classical statistical mechanics. To do so, one has to average the desired quantities over an ensemble. For a function $O({\mathbf x},{\mathbf p})$ of the phase space points such an average $\langle O({\mathbf x},{\mathbf p})\rangle$ is given explicitly through
\begin{equation}
\langle O({\mathbf x},{\mathbf p})\rangle=\dfrac{\int d^{3}x\, d^{3}p\, O({\bf x},{\bf p})\, e^{-\beta H({\bf x},{\bf p})}}{\int d^{3}x\, d^{3}p\, e^{-\beta H({\bf x},{\bf p})}},
\end{equation}
where $H({\bf x},{\bf p})$ is again the classical Hamiltonian.

In this context, the classical virial theorem states that for $N$ particles in thermal equilibrium at temperature $T$ in a potential $V$, the classical ensemble averaged potential energy $\left\langle V\right\rangle $ and translational kinetic energy $\left\langle K\right\rangle $ of a particle in the system are related according to \cite{greiner-st-mec}:
\begin{equation}
\left\langle K_{i}\right\rangle =\dfrac{1}{2}\left\langle \dfrac{\partial V}{\partial x_{i}}x_{i}\right\rangle =\dfrac{1}{2}k_{B}T,\label{eq:virial-1}
\end{equation}
where $x_{i}$ are the components of the position vector of a particle
and $K=\sum_{i}K_{i}$. For a cylinder symmetrical potential of the form $V=V(r,z)$, one has
\begin{eqnarray}
\hspace{-1cm}\left\langle K_{x}\right\rangle +\left\langle K_{y}\right\rangle  & = & \dfrac{1}{2}\left\langle r_{\perp}\dfrac{\partial V}{\partial r_{\perp}}\right\rangle,\quad 
\left\langle K_{z}\right\rangle  = \dfrac{1}{2}\left\langle z\dfrac{\partial V}{\partial z}\right\rangle .\label{eq:virial-2}
\end{eqnarray}
The anharmonic potential, as given in \eqref{eq:pot-2}, can be decomposed according to
\begin{equation}
V(r_{\perp},z)=V_{r_{\perp}}(r_{\perp})+V_{z}(z)
\end{equation}
with the terms
\begin{eqnarray}
V_{r_{\perp}}(r_{\perp}) & = & \dfrac{\epsilon_{z}}{2}\left[\lambda^{2}\eta\dfrac{r_{\perp}^{2}}{a_{z}^{2}}+\dfrac{\kappa}{2}\dfrac{r_{\perp}^{4}}{a_{z}^{4}}\right],\quad
V_{z}(z) = \dfrac{\epsilon_{z}z^{2}}{2a_{z}^{2}}.\label{eq:pot-3b}\end{eqnarray}
Using \eqref{eq:virial-2}, we obtain for the $z$-degrees of freedom the familiar harmonic potential result
\begin{equation}
\left\langle K_{z}\right\rangle =\left\langle V_{z}\right\rangle .
\end{equation}
For the radial degrees of freedom we find correspondingly
\begin{equation}
\left\langle K_{x}\right\rangle +\left\langle K_{y}\right\rangle =2\left\langle V_{r_{\perp}}\right\rangle -\lambda^{2}\eta\dfrac{\epsilon_{z}}{2}\dfrac{\left\langle r_{\perp}^{2}\right\rangle }{a_{z}^{2}}.
\end{equation}

The total internal energy $U$ of a system of $N$ non-interacting
particles in the potential $V$ is then given by
\begin{equation}
U=N(\left\langle K\right\rangle +\left\langle V\right\rangle )=\dfrac{5}{2}Nk_{B}T+\dfrac{N\lambda^{2}\eta\epsilon_{z}}{4}\dfrac{\left\langle r_{\perp}^{2}\right\rangle }{a_{z}^{2}}.\label{eq:U-virial}
\end{equation}

Before we consider the general result, let us make some remarks about Eq.~\eqref{eq:U-virial}. By using the radial symmetry of the problem, one can express the average square radius through
\begin{equation}
\langle r_{\perp}^{2}\rangle = \frac{m_{3}}{m_{1}},
\label{r_perp_sq_aver}
\end{equation}
with the momenta
\begin{equation}
m_{i}=\int_{0}^{\infty}\rho^{i}{\mathrm d}\rho e^{-\beta\left[\dfrac{M\omega_{\perp}^{2}\eta\rho^{2}}{2} + \dfrac{\kappa\rho^{4}}{4} \right]}.
\end{equation}

With the help of Ref. \cite{gradshteyn-tisp}, the limiting cases can be analytically explored. In the case of a pure quartic trap ($\eta=0$), the average radius is given by
\begin{equation}
\langle r_{\perp}^{2}\rangle^{\rm QU} = \frac{2}{\sqrt{\pi\beta\kappa}},
\label{r_perp_sq_aver_quar}
\end{equation}
showing, as expected, that $\kappa$ sets up the important length scale.

For the case of a pure harmonic trap, i. e. $\kappa=0$, one has
\begin{equation}
\langle r_{\perp}^{2}\rangle^{\rm HO} = \frac{2}{\beta M\omega_{\perp}^{2}\eta}.
\label{r_perp_sq_aver_harm}
\end{equation}

From \eqref{r_perp_sq_aver_harm}, we conclude that the average position of the particles diverges as $\eta$ approaches $0$. Nevertheless, the energy per particle, as given from \eqref{eq:U-virial}, remains $3k_{B}T$.

In general we obtain
\begin{eqnarray}
\langle r_{\perp}^{2}\rangle & = & \dfrac{2a_{z}^{2}}{(\kappa\beta\epsilon_{z})^{\frac{1}{2}}} \left\{-\sigma(\eta)(\beta \Delta E)^{\frac{1}{2}} + \right.\nonumber\\
& &\left. \dfrac{e^{-{\beta}{\Delta E}}} {\left[2\theta(-\eta) + \sigma(\eta)Q(\frac{5}{2},{\beta}{\Delta E})\right]\sqrt{\pi}}\right\}.
\end{eqnarray}
Substituting this into \eqref{eq:U-virial} we recover \eqref{eq:HighT-U}.
We, thus, see that the classical statistical treatment of the system
is equivalent to the high-temperature expansion which we have developed
by keeping only the leading term in \eqref{eq:HighT-Inc}.

\section{Conclusions and Outlook}
\label{sec:6}
We have considered non-interacting spin-polarized Fermi particles in an anharmonic trap in the presence of a centrifugal potential due to rotation of the system. Applying a semiclassical treatment to the density of states, we have worked out the temperature dependence of several thermodynamical properties. At absolute zero we have studied the density of particles and the momentum distribution for an anharmonically trapped system. The analysis of the particle density shows that a well defined geometry characterizes the ground state and has revealed a tendency for the particles to move away from the center. An analytical expression was obtained for the value of the rotation frequency $\Omega_{\rm DO}$ at which a hole is created at the center of the trap, giving rise to a regime where the gas is `donut'-shaped. The frequency $\Omega_{\rm DO}$ was shown to depart from $\omega_{\perp}$, which characterizes the stability limit for harmonic confinements, due to the interplay between the fermionic statistical interaction and the anharmonicity, in close analogy to repulsively interacting bosons. At finite temperature, expressions for the chemical potential, the internal energy, and the heat capacity were given. These were followed by discussions of their dependence on both anharmonicity and rotation frequency. At high-temperature, a virial theorem approach was presented and compared to previous results as well as shown to be particularly useful in the intuitive limits of pure harmonic and quartic radial confinements.

The most natural extension of this work would be the inclusion of interactions. In the low-temperature regime, short-range interactions are strongly suppressed for fully polarized fermions. To overcome this problem, one could include other Zeeman states by using an optical trap \cite{tl.ho-prl81}, yielding a spinor Fermi gas. Issues as (anti-)ferro\-mag\-net\-ism and magnetization dependence of the different quantities \cite{unitary} could be addressed in the presence of the additional quartic confinement. As for long-range interactions, no further Zeeman states are necessarily required. Due to its anisotropy, the dipole-dipole interaction is particularly interesting. Atoms with non-zero nuclear spin are possible constituents for systems with large magnetic dipole moment $m$. Degenerate stable, for $^{53}$Cr ($m=6\mu_{B}$) \cite{chromium53}, and meta-stable states, for $^{173}$Yb ($m=3\mu_{B}$) \cite{ytterbium}, are believed to be achievable. To date, only studies in two-dimensional systems consisting of a few atoms have already been performed \cite{dipole_2D}, so that a three-dimensional many-body analog could be thought of.

A further, and even simpler extension of the present work could be achieved by investigating the two-di\-men\-sional limit ($\omega_{z}\gg\omega_{\perp}$) of the present system. By an appropriate tuning of the quartic constant, a quantum Hall regime could be realized making possible the observation of underlying Landau-levels through a plateau structure in the particle density when $\Omega\approx\omega_{\perp}$ \cite{PhysRevLett.85.4648}. For higher rotation frequencies, one would expect this structure to disappear and a hole to be created at a frequency value analogous to $\Omega_{\rm DO}$.

\subsection*{Acknowledgements}

We acknowledge financial support from the German Academic Exchange Service (DAAD) and from the German Research Foundation (DFG) within the Collaborative Research Center SFB/TR12 Symmetries and Universalities in Mesoscopic Systems.

%

\end{document}